\documentclass[10pt]{article}

\usepackage{epsfig}

\def\beq{\begin{equation}}  

\def\eeq{\end{equation}}

\newcommand{\eq}{\begin{equation}}

\newcommand{\eqx}{\end{equation}}

\newcommand{\eqn}{\begin{eqnarray}}

\newcommand{\eqnx}{\end{eqnarray}}

\newcommand{\rr}[4]{#1, {\it #2 \/}{\bf #3} #4}

\begin{document}
 
\title{Popping out the Higgs boson off vacuum at Tevatron and LHC}
\author{M. Boonekamp\thanks{CERN, CH-1211, Geneva 23,
Switzerland and CEA/DSM/DAPNIA/SPP, CE-Saclay, F-91191
Gif-sur-Yvette Cedex, France}, R. Peschanski\thanks{%
CEA/DSM/SPhT,  Unit\'e de recherche associ\'ee au CNRS, CE-Saclay, F-91191 
Gif-sur-Yvette Cedex,
France}, C. Royon\thanks{%
CEA/DSM/DAPNIA/SPP, F-91191 
Gif-sur-Yvette Cedex,
France}}
\maketitle

\begin{abstract}
In the prospect of diffractive Higgs production at the LHC collider, we give 
an extensive study of Higgs boson, dijet, diphoton and dilepton production at 
hadronic colliders via diffraction at both hadron vertices. Our model, based 
on non factorizable Pomeron exchange, describes well the observed dijet rate 
observed at Tevatron Run I. Taking the absolute  normalization from data, our 
predictions  are given for diffractive processes 
at Tevatron and LHC. Stringent  tests of our model and of its parameters using 
data being taken now at Tevatron Run II are suggested. These measurements will 
also allow to discriminate between various models and finally to give precise 
predictions on diffractive Higgs boson production cross-section at the LHC. 
\end{abstract}

\section{Double Diffractive Hard Production through non-factorizable Vacuum
Exchange}

The discovery of the Higgs boson is one of the main goals of searches at 
the present and next hadronic colliders, the Tevatron
and the LHC. The standard, non-diffractive production mechanisms are being 
studied extensively. The main decay modes of a low-mass Higgs boson are 
$b$-quark 
pairs and $\tau$-leptons, which are difficult to extract from the
standard model background processes. A promising channel is the $\gamma\gamma$ 
decay mode; however, due to the small branching fraction (about 10$^{-3}$), a 
luminosity of the order of 50 fb$^{-1}$ is needed to establish the existence of 
the Higgs boson in the mass range below 135 GeV. It is thus important to 
investigate other, complementary ways to produce and detect the Higgs boson 
in this mass range.

One promising production mode, the {\it exclusive} double diffractive 
production, 
was proposed some time ago 
in Ref.~\cite{bi90,bjorken}. In this case, the Higgs boson is diffractively 
produced 
in the central region resulting in a final state composed of the two protons 
scattered 
at very small angles
and detected in the roman pot detectors, the decay products of the Higgs boson
in the main detector, and nothing else. It is thus a very clean signal. The 
kinematic constraints coming
from the proton detection in the roman pot detectors allow a very precise 
determination
of the Higgs boson mass \cite{albrow}, hence improving the signal 
to background ratio. Contrary to non-diffractive production, the main Higgs 
boson
decay modes, like $b \bar{b}$ or $\tau\tau$ are thus promising channels. 
However, the 
exclusive cross-sections may be very low, and thus put strong limitations to the 
potentialities of double diffractive production.

Recently, we have studied the possibility of producing the Higgs boson 
together with other particles, {\it i.e.} {\it inclusive} diffractive production
\cite{us}. The expected cross-sections are increased compared to exclusive 
production, and the model can be ``calibrated" using the diffractive dijet 
production 
measured by the CDF Collaboration at Tevatron run I \cite{CDF0}. The 
experimental result, and in
particular the dijet mass fraction spectrum, shows that hadronic activity in 
the central region (coming from hard QCD radiation, and from soft Pomeron 
remnants) 
needs to be accounted for, in order to describe these data. We have proposed a 
model for double diffractive production of heavy ``objects'', based on a 
non-factorizable 
Pomeron model and able to describe the observed features of dijet 
production data in a qualitative way. Normalizing our raw predictions to the CDF 
measurement 
allows to make quantitative predictions for the Tevatron and the LHC, given 
specific 
assumptions for the model parameters. 
We have also shown \cite{usbis} that it is possible to reconstruct precisely
the mass of the Higgs boson if  both  protons in the
final state can be detected with roman pot detectors and if the Pomeron remnants 
can be 
measured in the forward region with sufficient resolution.

At the Tevatron collider, the Higgs boson production 
cross-section is severely limited by the small available phase space. At the 
LHC,
since the beam energy is much higher, diffractive Higgs boson production
has larger cross-section, and might thus be an interesting channel, as we shall 
discuss 
further on. On the other hand, even if double diffractive Higgs boson
production at the Tevatron is probably too small by itself, it is however 
possible 
to verify and constrain models of double diffraction at the Tevatron Run II, 
through 
the study of difermion production.

In this paper, we wish to provide an extensive study of our model, including 
predictions for dijet, diphoton and dilepton production and their ratios for 
both the 
Tevatron run II and the LHC. We will discuss possible values of its 
characteristic 
parameters, and propose means to discriminate between the existing models at the 
Tevatron.  
In turn, this study will help to formulate more precise predictions for double 
diffractive Higgs boson production at the LHC in the near future.
 
The plan of our study is the following. In Section \ref{sectII}, we will recall 
and discuss the original exclusive model of Bialas and Landshoff \cite{bi90}, 
and our 
extension to inclusive production. In Section \ref{modpred} predictions are 
given for dijet, 
diphoton, dilepton and Higgs boson cross-sections. Dijet results 
for the Tevatron run I, using a fast simulation of detector effects, are in good 
agreement with available data and are used to normalize our predictions. A 
discussion of 
Pomeron remnants and 
their possible detection is also given. In Section \ref{parameter}, ways to 
verify our model and 
determine its parameters using Tevatron data are proposed. The possibility 
of distinguishing our model from the existing models of double diffraction is 
discussed 
in Section \ref{sectV}. Section \ref{sectVI} summarizes the interest of 
diffractive Higgs boson 
production compared to standard production. Finally, Section \ref{sectVII} gives 
conclusions and an 
outlook on the promising future studies on hard diffraction production at the
Tevatron and the LHC.

\section{Inclusive {\it vs.} 
exclusive production}
\label{sectII}

\subsection{Exclusive production}
\label{formulation}

\begin{figure}
\begin{center}
\epsfig{figure=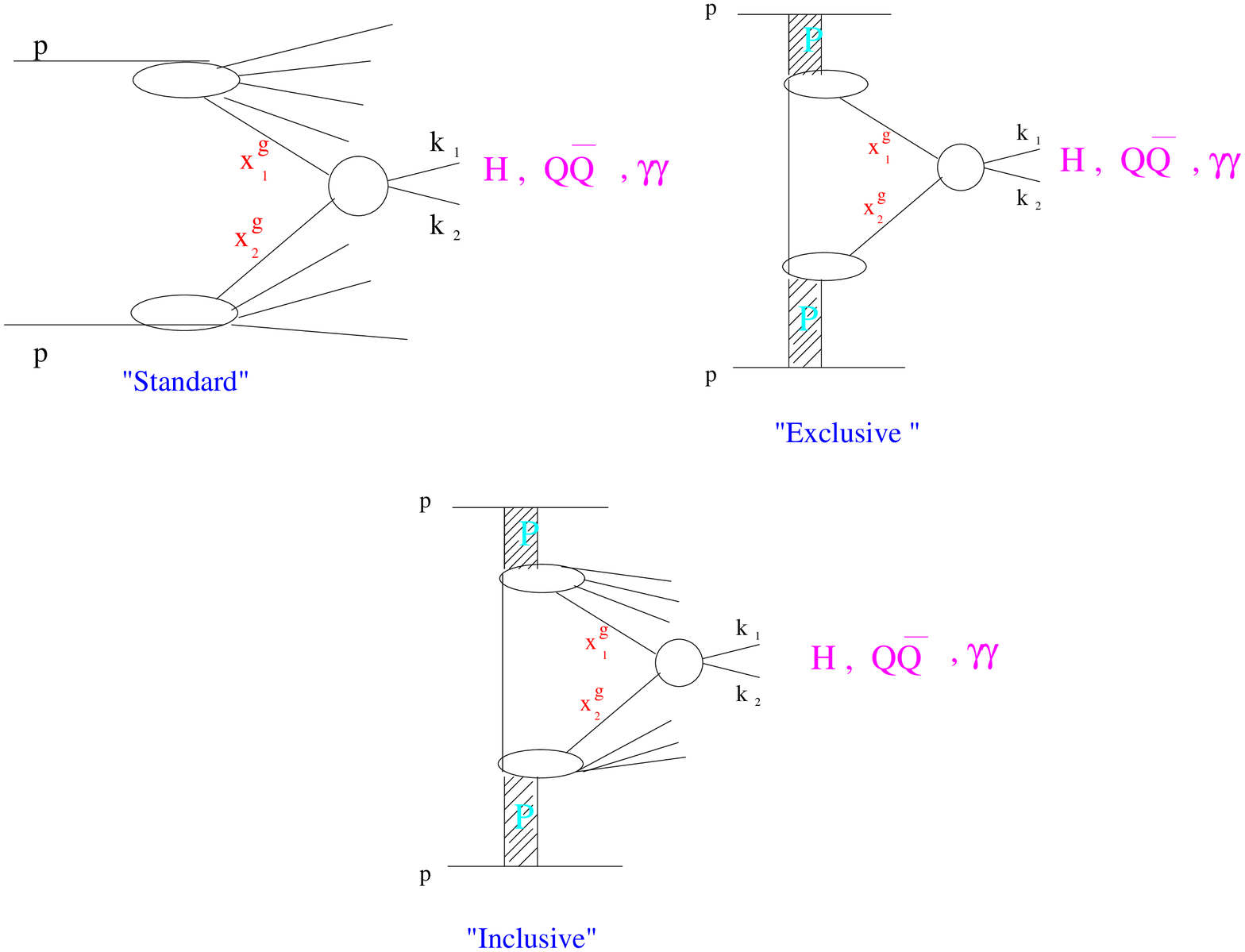,height=4.in}
\end{center}
\caption{{\it Double diffractive  vs. standard gluon fusion production 
schemes.} 
$x_i^g$ are the momentum fractions of the fusing gluons, $k_i$ are the 
transverse 2-momenta of the outgoing difermion in the central 
rapidity region. The ``Standard'' Higgs boson production is displayed for 
reference. ``Exclusive'' and ``inclusive'' double 
diffraction are represented in the framework of the non factorizable Pomeron 
model (see formulae (\ref{exclusif},\ref{dinclujj}-\ref{dinclugg}) and text for 
the 
complete kinematical notations). The hatched region represents the diffractive 
interactions at both 
(anti)proton vertices, while the vertical thin line is for the soft gluon 
exchange in the model. Note that $x_i^g \equiv 1$ in the ``exclusive'' case.}
\label{diag}
\end{figure}

Let us first introduce the original model of \cite{bi90} describing exclusive
Higgs boson and $q \bar q$ production in double diffractive production 
(noted DPE \footnote{We keep the now standard notation DPE ({\it i.e.} Double 
Pomeron 
Exchange) but the model we describe, together with its extension to 
inclusive production, considers a non-factorizable soft 
pomeron exchange with one common exchanged gluon, while  
a factorizable double Pomeron mechanism implies two different pairs of gluons 
(see Section \ref{parameter}).} in the following). This process is depicted in 
Fig.~\ref{diag}, 
``Exclusive'' case. 

In \cite{bi90}, the diffractive mechanism is based on two-gluon exchange between 
the 
two incoming protons. The soft Pomeron is seen as a  pair of gluons 
non-perturbatively
coupled  to the proton. One of the gluons is then coupled 
perturbatively to the hard process (either the Higgs boson, or the $q \bar q$ 
pair, see Fig.~\ref{diag}), while the other one plays the r\^ole of a soft 
screening of 
color, 
allowing for diffraction to occur. This soft character  requires the 
phenomenological introduction of a distinctive non-perturbative gluon 
propagator 
\cite{propa} whose 
parameters are constrained by the  description of total 
cross-sections using the same formalism. 

The hard gluons, carrying all 
remaining momentum ($x_1^g=x_2^g=1$), fuse to produce the heavy object (Higgs 
boson and diquarks in the original model).
The corresponding cross-sections for $q \bar{q}$ and Higgs boson production 
read:

\begin{eqnarray}
d\sigma_{q \bar{q}}^{exc}(s) &=& C_{q \bar{q}}^{exc} 
\left(\frac{s}{M_{q \bar{q}}^{2}}\right)^{2\epsilon}
                    \delta^{(2)}\left( \sum_{i=1,2} (v_{i} + k_{i}) \right)
                    \prod_{i=1,2} \left\{ d^{2}v_{i} d^{2}k_{i} d\eta_{i}\ 
                    \xi_{i}^{2\alpha'v_{i}^{2}}\! \exp(-2\lambda_{JJ} 
v_{i}^{2})
                    \right\} {\sigma^{exc}_{q \bar{q}}}; \nonumber \\
d\sigma_{H}^{exc}(s) &=& C_{H}\left(\frac{s}{M_{H}^{2}}\right)^{2\epsilon}
                    \delta\left(\xi_{1}\xi_{2}-\frac{M_{H}^{2}}{s}\right)
                    \prod_{i=1,2} \left\{ d^{2}v_{i} 
\frac{d\xi_{i}}{1-\xi_{i}} 
                   \ \xi_{i}^{2\alpha'v_{i}^{2}} \exp(-2\lambda_{H} v_{i}^{2})
                    \right\}\ .
                    \label{exclusif}
\end{eqnarray}

The variables $v_{i}$ and $k_{i}$ respectively denote the transverse
momenta of the outgoing protons and quarks, $\xi_{i}$ are the proton
fractional momentum losses\footnote{The $\xi_i$ are given by $\xi_{1,2}\sqrt{s}= m_{T1}\exp{\pm\eta_1}+m_{T2}\exp{\pm\eta_2}$, $m_{Ti}=k_i^2 + m_q^2$,
and restricted to be smaller than $10\%$. This value for 
$\xi \le 0.1$ comes from the usual cut on rapidity gap to be $\ge 2.$},
and $\eta_{i}$ are the quark rapidities. $\sigma_{H}$ is the
gluon-initiated Higgs boson production cross-section while ${\bar 
\sigma_{q \bar{q}}}^{exc}$ is the hard $q \bar q$ 
production cross-section in the 
exclusive case. Indeed, in this exclusive process submitted to the $J_{Z}=0$ 
constraint \cite{pumplin} \footnote{Note that the exclusive model of formula (1) 
is related 
to the soft pomeron exchange with a low value of the intercept, by contrast with 
the
model of Ref. \cite{khoze}.} (a helicity selection rule for the production of a 
scalar
system from the vacuum channel), the
$q \bar q$ differential cross-section writes\footnote{Here, as in the rest of 
the paper, the hard differential cross-sections 
${\sigma}$ are normalized to the  usual $d\sigma/dt$ expressions through $
{\sigma} = \frac{\pi}{24} \frac{d\sigma}{dt}.$}, apart from normalizations
included in $C_{q \bar{q}}$:
\begin{equation}
{\sigma}_{q \bar{q}}^{exc} \equiv \frac{\pi}{24} \frac{d\sigma}{dt} = 
\frac{\rho(1-\rho)}{m_{T1}^{2}m_{T2}^{2}} \,\, , 
\,\,
\rho = \frac{4m_{Q}^{2}}{M_{q \bar{q}}^{2}}\ .
\label{exc}
\end{equation}

In the model, 
the non-perturbative input  is naturally related to the soft Pomeron 
trajectory 
taken 
from the standard Donnachie-Landshoff   parametrization
\cite{pom}, namely $\alpha(t) = 1 + \epsilon + \alpha't$, with
$\epsilon \approx 0.08$ and $\alpha' \approx 0.25 \mathrm{GeV^{-2}}$.

There are other parameters to be fixed, coming from the non-perturbative gluon 
propagators. Phenomenological  contraints are obtained from the physical 
values of the total cross-sections, leading to four unknown parameters in 
formula (\ref{exclusif}), namely the normalizations $C_{q \bar{q}},\ C_{H}$ and 
the slopes in  momentum transfer $\lambda_{q \bar{q}},\lambda_{H}.$ 
At this point of our study, the  slopes $\lambda_{q \bar{q}},\lambda_{H}$ are 
kept 
as in  the original papers \cite{bi90} \footnote{They are constrained by using 
(an approximate 
parametrization of) the nucleon form factor as the Pomeron coupling to proton 
\cite{propa}.}.

The problem of the normalization constants $C_{q \bar{q}},\ C_{H}$ requires 
more care, since the evaluation of cross-sections is the main subject of the 
paper. 
All constants of the problem (non-perturbative
proton-gluon coupling, normalization of the non-perturbative gluon
propagators, color factors, perturbative Higgs boson and $q \bar q$
production vertices) are contained in the normalizations $C_{q \bar{q}}$ and
$C_{H}$. The non-perturbative factors are poorly determined by the
current data, implying large uncertainties on production cross-section
predictions. In particular, even if every other parameter is fixed, the 
non-perturbative proton-gluon coupling $G$ remains undetermined and is 
arbitrarily fixed 
to $G^2/4\pi =1$ in the original publications \cite{bi90}. However, and it is an 
important aspect of our  model studies,  the ratio $C_{H}/C_{q \bar{q}}$ is
well defined, and independent of everything else than the color
structure of the process. This means that the {\it ratio} of Higgs boson to 
diquark production is fixed by this factor. Given the expressions 
of $C_{H}$ and $C_{q \bar{q}}$ \cite{bi90}, one finds :
\begin{equation}
\left(\frac{C_{H}}{C_{q \bar{q}}}\right) = \frac {\sqrt2 {\cal G}_F}{3\pi}\ ,
\label{ratio} 
\end{equation}
where enter only in  (\ref {ratio}) the Fermi constant ${\cal G}_F$, and the 
ratio of color factors to produce either a color singlet  $q \bar q$  or the 
scalar Higgs 
boson in the non-factorizable Pomeron model (see for details, the second 
reference of \cite{bi90}). This feature will remain valid in our model of 
inclusive 
production, suggesting that the known, large cross-section dijet 
production process can be used to calibrate Higgs boson production.

Also note that, as can be seen in  (\ref{exc}), the exclusive  production rate 
for a
given quark flavour is proportional to its mass squared, so that light quark
production is expected to be negligible, reflecting in another way the 
$J_{Z}=0$ 
constraint \cite{pumplin}.

\subsection{Inclusive production}
\label{inclformulation}

The inclusive mechanism is described in the third graph of Fig.~\ref{diag}. The 
idea is 
to take into account that a Pomeron is a composite system, made itself from 
quarks and 
gluons. In our model, we thus apply the concept of Pomeron structure functions 
to 
compute the inclusive diffractive Higgs boson cross-section. The H1 
measurement of the diffractive structure function \cite{H1pom} and the 
corresponding quark and gluon 
densities are used for this purpose. This implies the existence of
Pomeron remnants and QCD radiation, as is the case for the proton. This 
assumption comes
from {\it QCD factorisation} of hard processes. 
However, and this is also an important issue, we do not assume {\it Regge 
factorisation} at the proton vertices, {\it i.e.} we do not use the H1 Pomeron 
flux factors in 
the proton or antiproton.

Regge factorisation is known to be violated between HERA and the
Tevatron. Moreover, we want to use the same physical idea as in the 
exclusive model \cite{bi90}, namely that a non perturbative gluon exchange 
describes the 
soft interaction between the incident particles, as in Fig.~\ref{diag}. In 
practice, the 
Regge factorisation breaking appears in three ways in our model:

{\bf i)} We keep as in the original model of Ref \cite{bi90} the soft Pomeron
trajectory with an intercept value of 1.08.

{\bf ii)}  We normalize our
predictions to the CDF Run I measurements, allowing for factorisation breaking 
of the Pomeron flux factors in the normalisation between the HERA and hadron 
colliders \footnote{Indeed, recent results from a QCD fit to the diffractive 
structure 
function in H1 \cite{newman} show that the discrepancy between the gluonic 
content of
the Pomeron at HERA and Tevatron \cite{goulianos} appears mainly in
normalisation.}.

{\bf iii)} The color factor (\ref{ratio}) derives from the non-factorizable 
character of the model, since it stems from the gluon exchange between the 
incident hadrons. We will see later the difference between this and the 
factorizable
case.

The formulae for the inclusive production processes considered here follow. We 
have, 
for dijet production\footnote{We call ``dijets'' the produced quark and gluon 
pairs.}, 
considering only the dominant gluon-initiated hard processes:
\eject 
\begin{eqnarray}
d\sigma_{JJ}^{incl} = C_{JJ}\left(\frac {x^g_1x^g_2 s }{M_{J
J}^2}\right)^{2\epsilon}\!\! \! \! \delta ^{(2)}\! 
\left( \sum _{i=1,2}
v_i\!+\!k_i\right) \prod _{i=1,2} \!\!\left\{{d\xi_i}  {dx_i^g}
d^2v_i d^2k_i {\xi _i}^{2\alpha' v_i^2}\!
\exp \left(-2 v_i^2\lambda_{JJ}\right)\right\} \times\nonumber \\
\times\left\{{\sigma_{JJ}} G_P(x^g_1,\mu) G_P(x^g_2,\mu) \right\};
\label{dinclujj}
\end{eqnarray}

\noindent and for Higgs boson production:

\begin{eqnarray}
d\sigma_H^{incl} = C_{H}\left(\frac {x^g_1x^g_2 s
}{M_{H}^2}\right)^{2\epsilon} \!\!\delta \left(\xi _1 \xi
_2\!-\!\frac{M_{H}^2}{x^g_1x^g_2 s} \right) \!\!\prod _{i=1,2}
\left\{G_P(x^g_i,\mu)\ dx^g_i  d^2v_i\ \frac {d\xi _i}{1\!-\!\xi 
_i}\
{\xi _i}^{2\alpha' v_i^2}\ \exp \left(-2 
v_i^2\lambda_H\right)\right\};
\label{dincluH}
\end{eqnarray}

\noindent For the two following processes, the quark-initiated contribution can 
not be ignored.
We have, for dilepton production:

\begin{eqnarray}
d\sigma_{ll}^{incl} = C_{ll}\left(\frac {x^q_1x^q_2 s }{M_{l
l}^2}\right)^{2\epsilon}\!\! \! \!{\sigma_{q\bar{q} \rightarrow ll}}\delta 
^{(2)}\! 
\left( \sum _{i=1,2}
v_i\!+\!k_i\right) \!\!\prod _{i=1,2} \!\!\left\{Q_P(x^g_i,\mu)  {d\xi_i}  
{dx_i^g}
d^2v_i d^2k_i {\xi _i}^{2\alpha' v_i^2}\!
\exp \left(-2 v_i^2\lambda_{ll}\right)\right\};
\label{dinclull}
\end{eqnarray}

\noindent and for diphoton production:

\begin{eqnarray}
d\sigma_{\gamma 
\gamma}^{incl} = C_{ll}\left(\frac {x^g_1x^g_2 s }{M_{\gamma 
\gamma}^2}\right)^{2\epsilon}\!\! \! \! \delta ^{(2)}\! 
\left( \sum _{i=1,2}
v_i\!+\!k_i\right) \prod _{i=1,2} \!\!\left\{{d\xi 
_i}  {dx_i^g}
d^2v_i d^2k_i {\xi _i}^{2\alpha' v_i^2}\!
\exp \left(-2 v_i^2\lambda_{\gamma 
\gamma}\right)\right\} \times\nonumber \\
\times\left\{{\sigma_{gg \rightarrow \gamma \gamma}} G_P(x^g_1,\mu) 
G_P(x^g_2,\mu) + \sigma_{q\bar{q} \rightarrow \gamma \gamma} Q_P(x^g_1,\mu) 
Q_P(x^g_2,\mu)\right\}.
\label{dinclugg}
\end{eqnarray}

\noindent In the above, $x_i^g$ are
the Pomeron's momentum fractions carried by the gluons or quarks involved in the hard process, and the $G_P$ (resp. $Q_P$) are the Pomeron gluon (resp. quark) 
energy densities, $i.e.$ the parton density multiplied by $x_i^g $. We use as 
parametrizations of the Pomeron structure functions the fits to the diffractive 
HERA data 
performed in \cite{ba00}. The dijet cross-section\footnote{The formulae 
(\ref{inclusif}) are 
corrected for a factor $2$ error coming from a known misprint in the 
normalization of ${\sigma}_{ff}$ in \cite{co83}.} is now (summing 
over quark flavours $f$, and now including the contribution from gluon
jets):

\begin{eqnarray}
{\sigma}_{JJ} = \sum_{f} {\sigma}_{gg \rightarrow q\bar{q}}(\rho^{f}) + 108\ 
{\sigma}_{gg \rightarrow gg}(\rho^{g}) \,\, ;
\,\, \rho^{f} = \frac{4 m^{f}_{T1} m^{f}_{T2}}{M_{JJ}^{2}} \,\, \ ; 
\,\, \rho^{g} = \frac{4 p_{T1} p_{T2}}{M_{JJ}^{2}}; \nonumber \\
{\sigma}_{gg \rightarrow q\bar{q}} = \frac{\rho^{f}}{{m^{f}_{T1}}^{2}
{m^{f}_{T2}}^{2}}\left(1-\frac{\rho^{f}}{2}\right)\left(1-\frac{9\rho^{f}}{16}
\right) 
\,\, ; \,\,
{\sigma}_{gg \rightarrow gg} = \frac{1}{p_{T1}^{2} 
p_{T2}^{2}}\left(1-\frac{\rho^{g}}{4}\right)^{3};
\label{inclusif}
\end{eqnarray}

\noindent to be compared with (\ref{exc}). The above formulae are derived using 
\cite{kripfganz}, and the dilepton and diphoton cross-sections are taken from 
\cite{eichten,braaten}. The expressions for $\sigma_{gg \rightarrow q\bar{q}},
\sigma_{gg \rightarrow gg},\sigma_{q\bar{q} \rightarrow ll},$ 
and $\sigma_{gg \rightarrow \gamma\gamma},\sigma_{q\bar{q} \rightarrow 
\gamma\gamma}$ 
in terms of the Mandelstam variables are recalled in the Appendix.

In the inclusive case, contrary to the exclusive case,  dijet production is 
flavour democratic and thus the $\sum_{f}$ in (\ref{inclusif}) extends over all 
flavors except for the too massive top quark, due to kinematics. 
Note that the non perturbative parameters are kept the same as in the 
exclusive 
case. Indeed, the expressions (\ref{exclusif}) can be  recovered in the limit 
$G_P \rightarrow \delta(x-1)$, and substituting back equation (\ref{exc}) 
instead of 
(\ref{inclusif}) for the hard cross-section, which restricts 
to the $J_Z = 0$ component of the $q \bar q$
cross-section and reintroduces the flavor mass hierarchy. The 
normalization of $G_P$ is not determined by the HERA data (since it is mixed 
with the flux factors) but 
is  fixed in the model in order to lead to the same result for the 
energy-momentum sum 
rule
as for the exclusive case. It is interesting to note that 
the comparison with the observed dijet production rates will give the correct 
order of magnitude for the inclusive model, while the exclusive one leads to 
too 
important rates\footnote{One may think naively that this overestimation using 
parameters 
from soft hadronic cross-sections might be an argument against the non 
perturbative gluon  model. This argument does not hold against the inclusive 
predictions which gives already the correct order of magnitude.}, 
as discussed in the next section. Note that  the 
normalization cancels in the ratio
$\left(\frac{C_{H}}{C_{JJ}}\right)$, with the same value as in 
the exclusive case, see \ref{ratio}.

\section{Model predictions for DPE}
\label{modpred}

\subsection{DPE Dijets at the Tevatron Run I}

The CDF measurement of double diffractive dijet production \cite{CDF0} is
used as a verification of the validity of our approach (namely
concerning the inclusive picture we consider, and the application of
structure functions measured in electron-proton collisions in our
context). Once the model validity tested, the measured cross-section will  
allow us to fix (within experimental errors) the absolute 
normalization of the cross-sections. 

In the CDF measurement, an outgoing antiproton is measured on one side of
the detector, and the DPE nature of the events is ensured by requiring
a rapidity gap on the opposite side. The selection then requires at
least two jets satisfying a transverse energy criterion. The details of
the selections can be found in \cite{CDF0}. The measured cross-section is
$43.6 \pm 4.4 \mathrm{(stat)}\pm 21.6 \mathrm{(syst)}$ nb, with large error
bars.

Reproducing the experimental selections on the cross-section estimates, and 
keeping fixed all parameters as in the exclusive case
we obtain a raw prediction of 11.4 nb, i.e. a factor 3.8 smaller than
the measured mean value \footnote{At a
theoretical level, it is interesting to note that a mere reduction of  the 
non-perturbative
proton-gluon coupling (arbitrarily fixed in original papers \cite{bi90})
from $G^2/4\pi =1\to  1/2$ swallows the normalization factor.}. This 
prediction includes a fast simulation of
detector resolution effects, using {\tt SHW} \cite{shw}. Considering the
large uncertainties, this result is quite encouraging. Indeed, as mentioned 
above, the experimental errors are yet quite large. 

\begin{figure} [p]

\begin{center}
\epsfig{figure=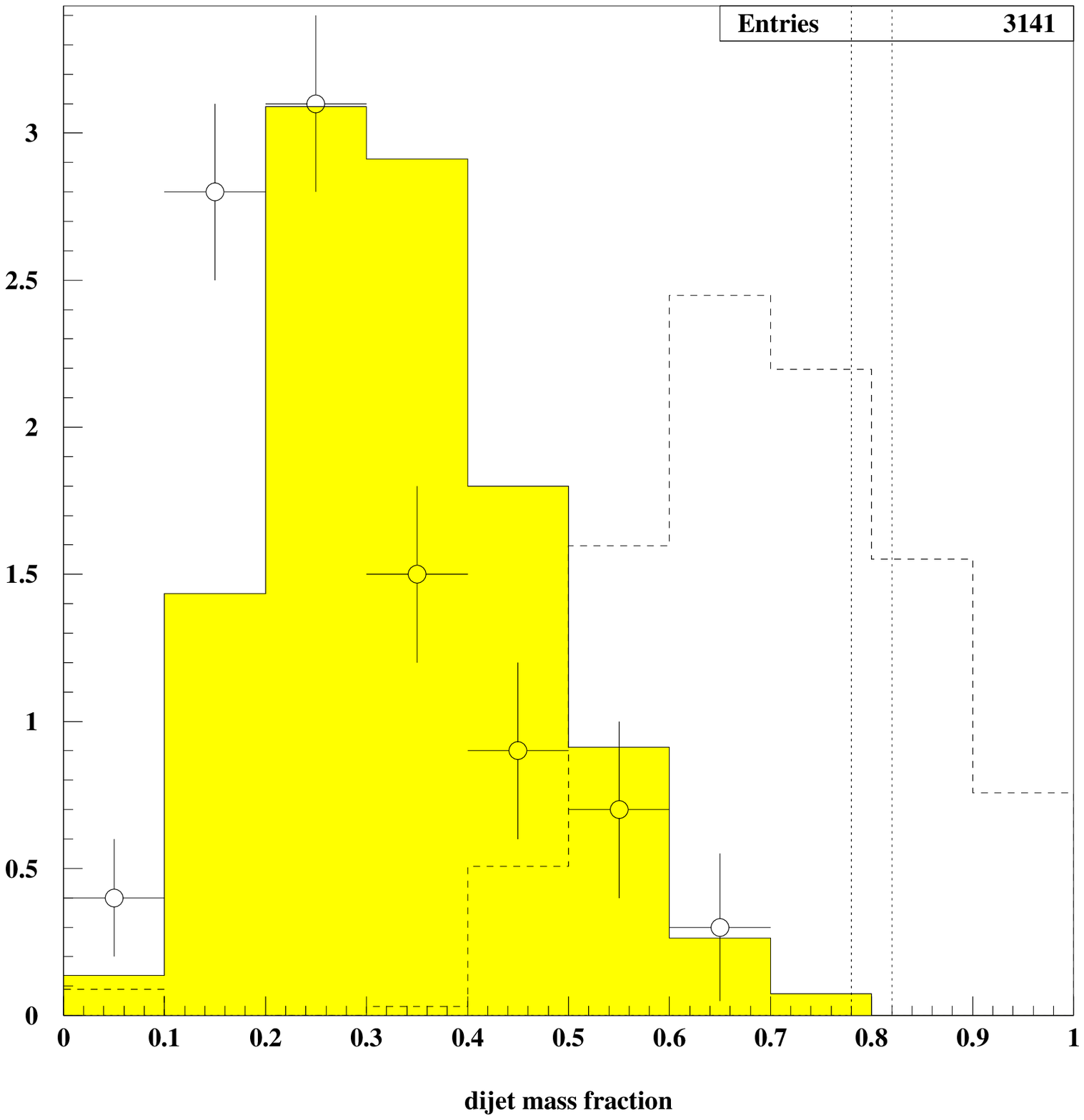,height=3.in}
\end{center}
\caption{Observed dijet mass fraction (CDF Run I), compared to our model 
prediction, using the Pomeron structure functions. The dashed histograms are for exclusive
production with and without ($cf.$ the peak at 0.8) detector simulation.}
\label{dijetmassfraction}

\begin{center}
\epsfig{figure=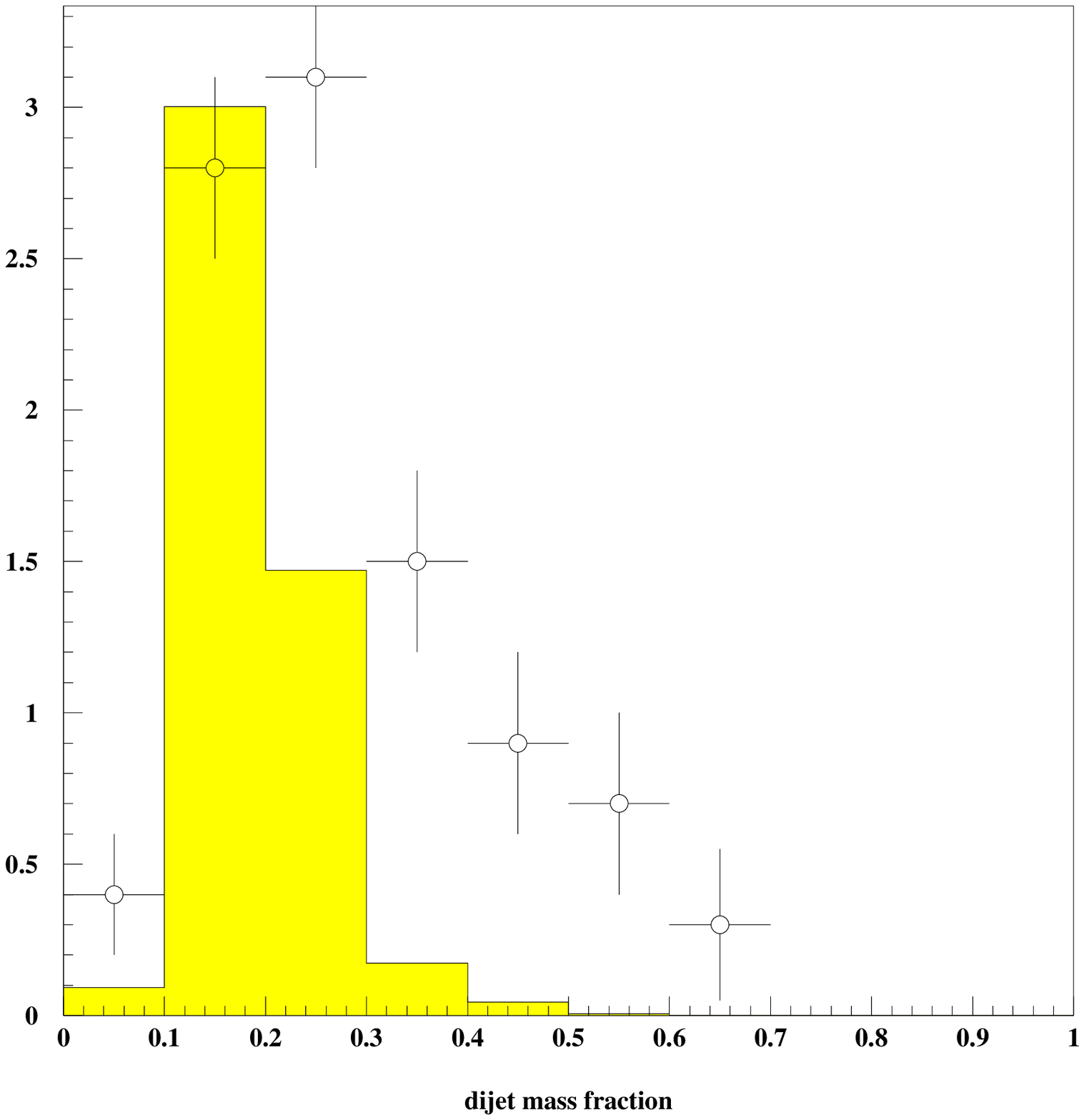,height=3.in}
\end{center}
\caption{Dijet mass fraction obtained using the gluon structure function from 
the proton.}
\label{protonsf}

\end{figure}

In order to verify the dynamics of the
model, it is interesting to consider the dijet mass fraction,
defined as the ratio of the mass measured in the central detector to
the missing mass to the outgoing proton and antiproton. For exclusive
events, this ratio is expected to be about 0.8 \cite{CDF0} given detector 
inefficiencies;
if inclusive events dominate, and the model is correct, one should observe a 
broad distribution below one, essentially given by the product of the Pomeron 
structure 
functions. Fig.~\ref{dijetmassfraction} displays the mass fraction as 
measured by CDF, and the prediction
of the present model \footnote{Recall that in our predictions, we only included
the gluon structure functions in the pomeron ($G_P$), neglecting the quark
structure function. This assumption is justified by the weakness of the quark
structure function (10 \%), and the large error bar on the
gluon structure function of about 50\% \cite{ba00}.}. 

Reasonable agreement is observed, suggesting that
the HERA Pomeron structure functions allow for a correct description of
inclusive DPE events. It is worthwhile to emphasize the strong influence 
of the behaviour of the (hard) gluon distribution of the Pomeron on the 
predicted mass fraction. 
In Fig.~\ref{protonsf}, we consider a ``proton-like'' ($i.e$, soft) gluon 
distribution 
in the Pomeron, leading to unsatisfactory results.

\subsection{Model predictions for Tevatron and LHC: dijets, diphotons and 
dileptons}

Assuming a global normalization as determined here and all other model
parameters as given above, one is  now in situation to give predictions 
for Higgs boson production at the Tevatron and the LHC. In a first stage,  we 
assume no
evolution of this normalization with energy. We will discuss later on  the 
influence of the various parameters on the predictions, analyze the 
(large) uncertainties which still affect the  results and, above all, what can 
be 
done in the future to handle them, both on experimental and theoretical 
grounds.
For this sake, after scaling our model to the CDF Tevatron run I predictions, 
we  give
predictions for various DPE processes which are relevant for Tevatron Run II 
and 
LHC
measurements.

\begin{figure}[p]
\begin{center}
\epsfig{figure=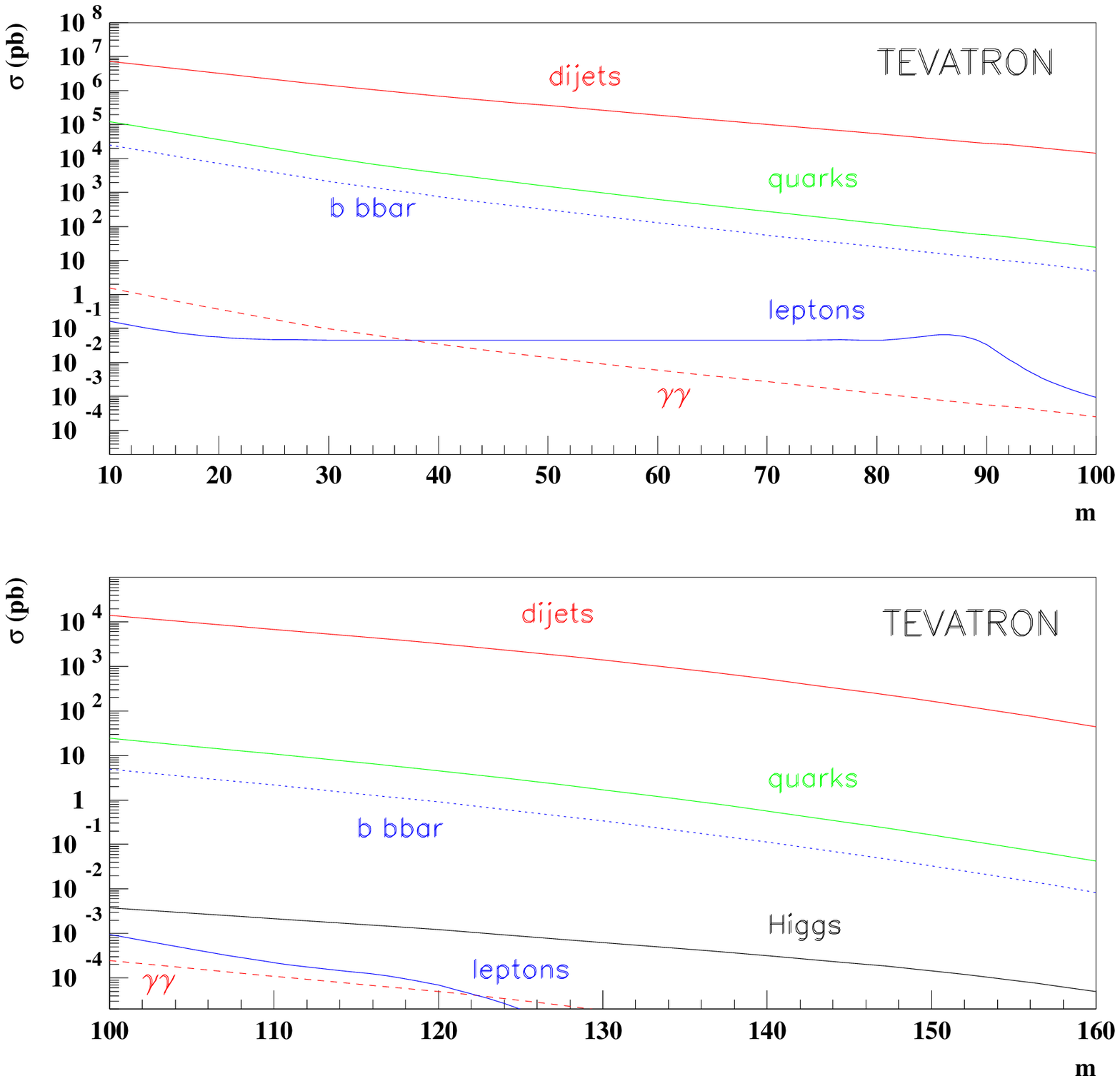,height=6.in}
\end{center}
\caption{Dijet, diquark, $b \bar{b}$, dilepton, and diphoton cross-sections (pb) 
at the Tevatron.
The cross-sections are given for a mass $m$ above the value on the abscissa, for
two different mass ranges. 
For comparison, we also display the Higgs boson cross-section
for $M_{Higgs}=m$.} 
\label{crosster}
\end{figure}

\begin{figure}[p]
\begin{center}
\epsfig{figure=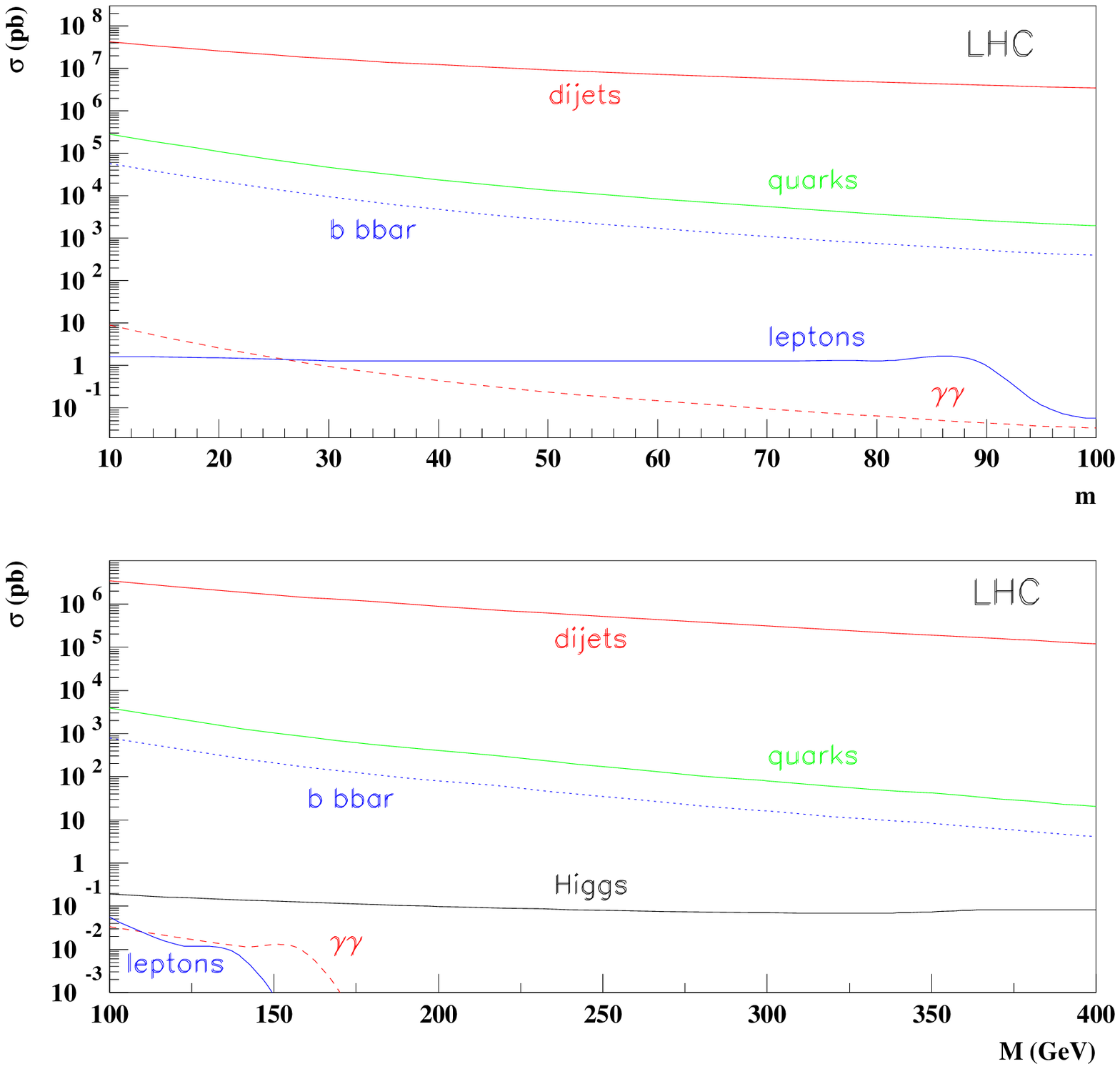,height=6.in}
\end{center}
\caption{Dijet, diquark, $b \bar{b}$, dilepton, and diphoton cross-sections (pb) 
at the LHC. The cross-sections are given for a mass $m$ above the value
on the abscissa, for two different mass ranges. For comparison, we also
display the Higgs boson cross-section for $M_{Higgs}=m$.} 
\label{crosslhcb}
\end{figure}

In Figs.~\ref{crosster} and \ref{crosslhcb}, we give the integrated cross 
sections
for dijets, quarks, $b \bar{b}$, dileptons and diphotons at generator level
for a mass higher than the
mass $m$ given on the abscissa for the Tevatron and the LHC.
Each figure is displayed for
two different mass ranges, namely between 10 and 100 GeV, and 100 and 160 GeV.

\begin{figure}[p]

\begin{center}
\begin{tabular}{cc}
\epsfig{figure=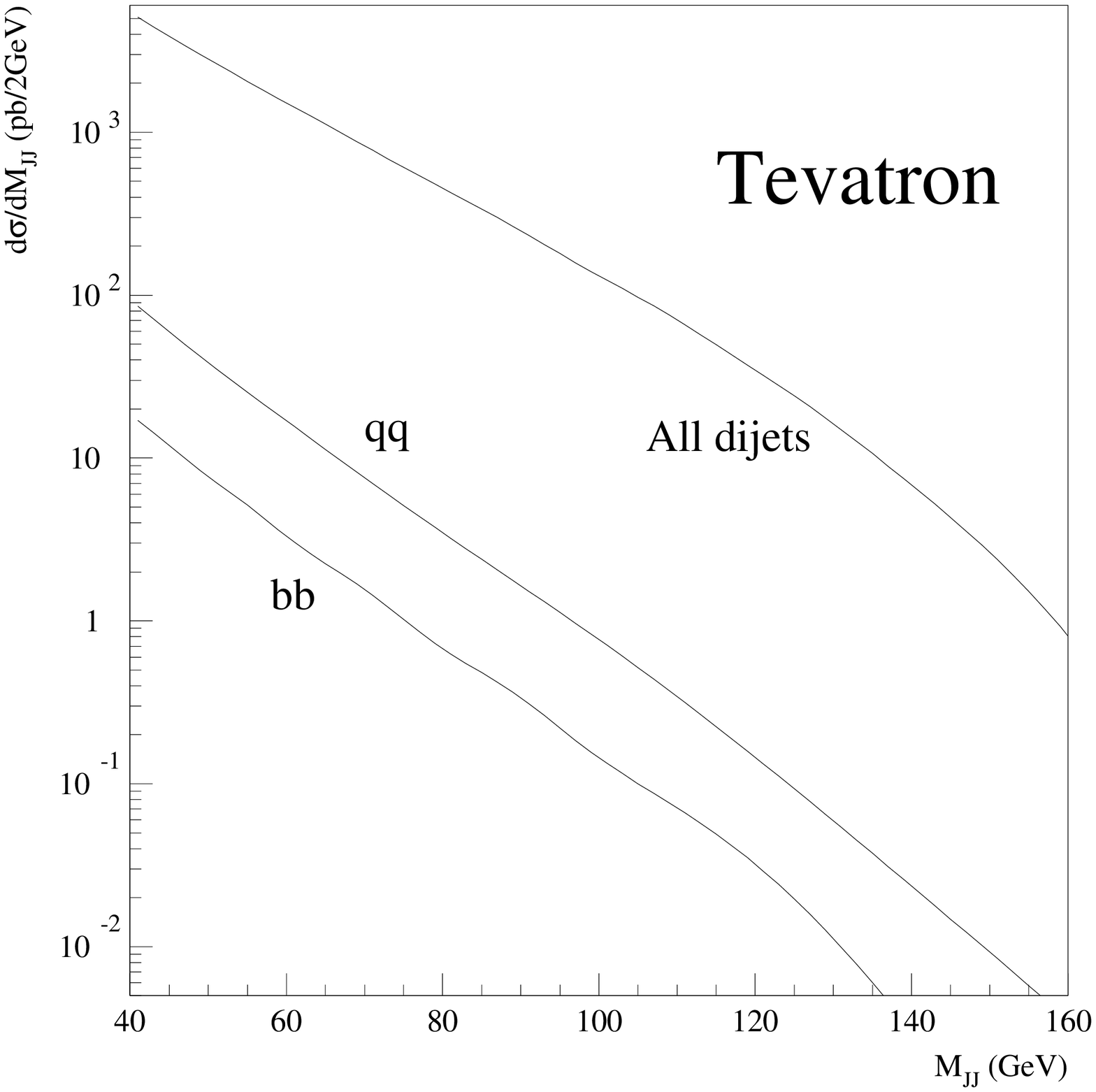,height=2.9in} &
\epsfig{figure=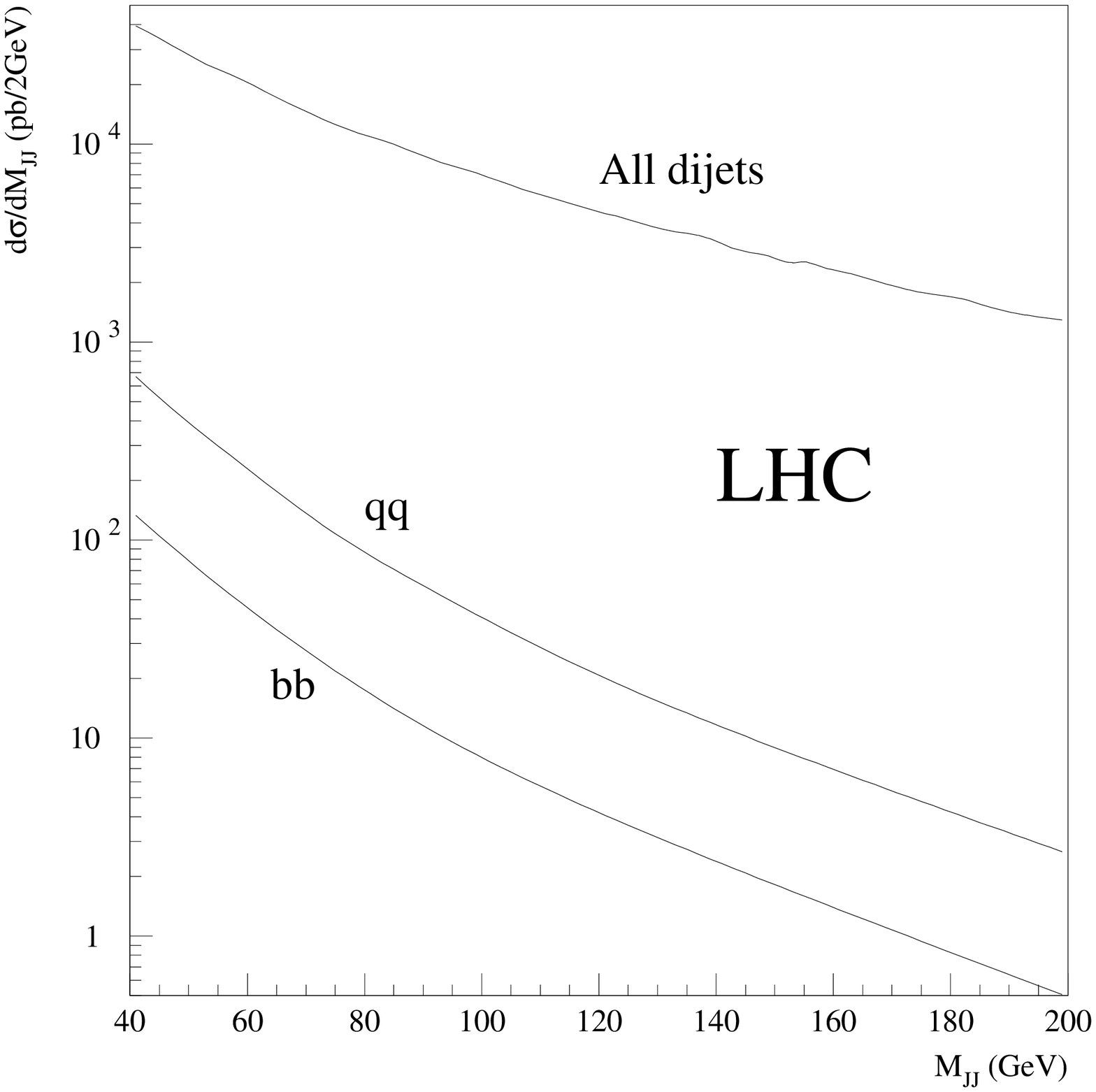,height=2.9in} \\
\end{tabular}
\end{center}
\caption{Differential dijet production cross-section (pb) at the Tevatron and 
the 
LHC. The transverse energy of the central jets satisfies $E_T > 10$ GeV, 
and their rapidity is limited to  $|y|<4$.}
\label{dijetXS}

\begin{center}
\begin{tabular}{cc}
\epsfig{figure=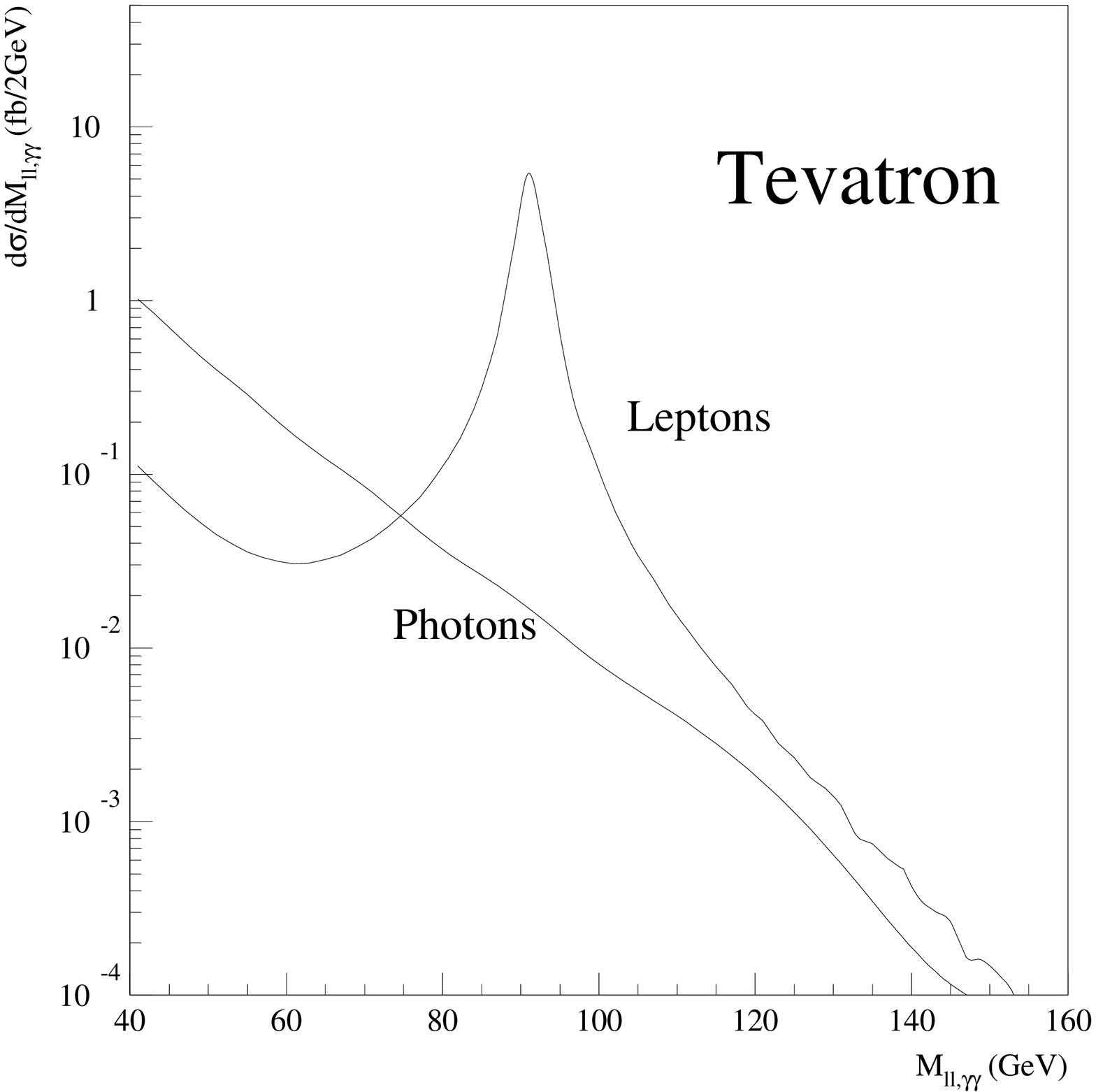,height=2.9in} &
\epsfig{figure=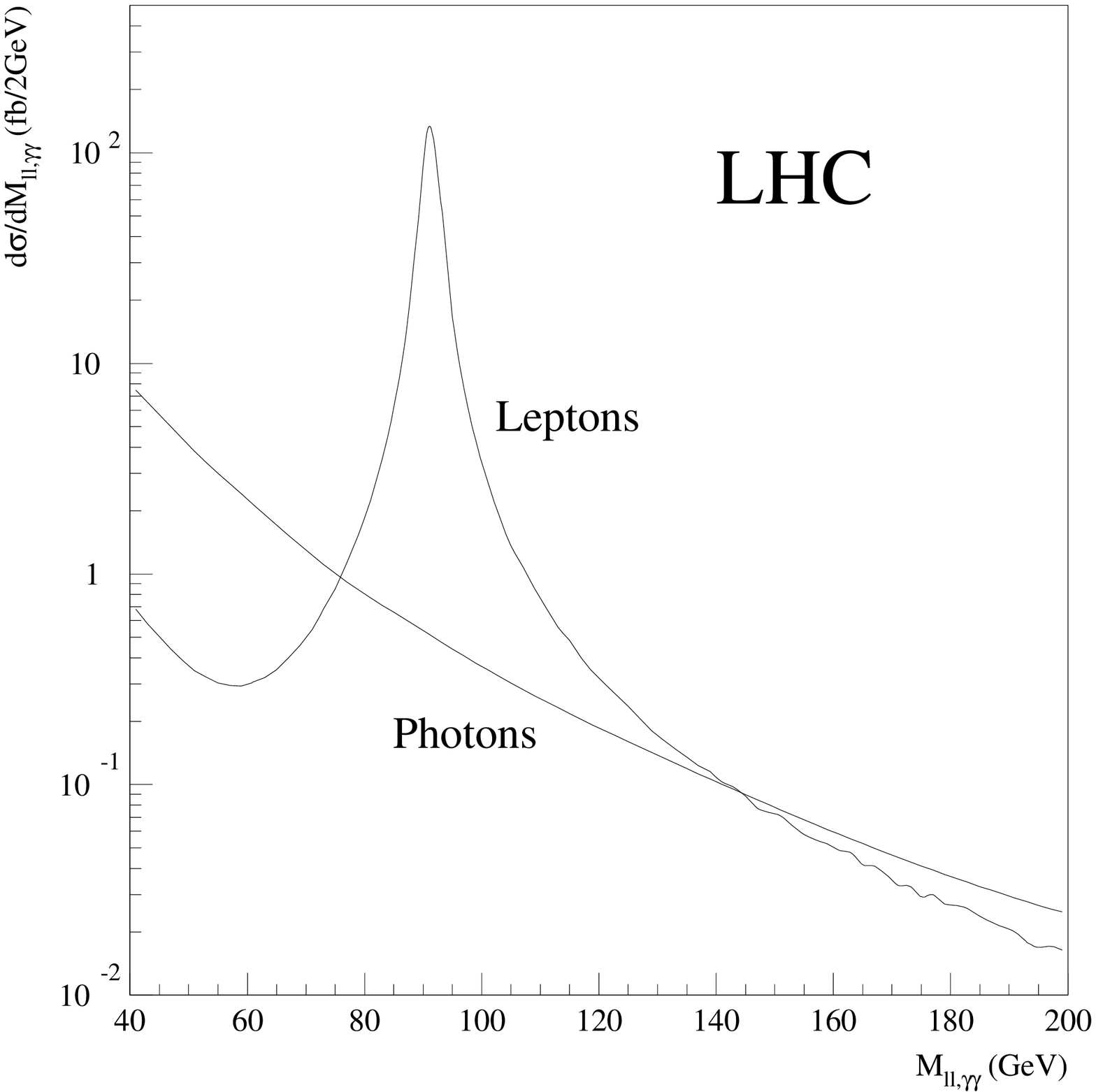,height=2.9in} \\
\end{tabular}
\end{center}
\caption{Differential diphoton and dilepton production cross-sections (fb) at 
the 
Tevatron and the LHC. The 
dilepton cross-section corresponds to a single lepton flavour. The 
transverse energy of 
the central particles satisfies $E_T > 10$ GeV, and their rapidity is 
limited to  $|y|<4$.}
\label{photonXS}

\end{figure}

The differential cross-sections for dijet, dilepton and diphoton production
are also given in Fig.~\ref{dijetXS} and \ref{photonXS}. The dijet cross-section 
is 
dominated by the gluon contribution. The quark jet contribution amounts 1\% at 
small masses,
and goes down to 0.1\% at higher masses, both at the Tevatron and the LHC. 
The $b \bar{b}$ contribution, which is about 20\% of the total quark 
contribution, 
represents only about 0.02\% of the dijet
yield. The diphoton cross-section is small (4 orders of magnitude below the
$b \bar{b}$ cross-section both at Tevatron and LHC) 
but still measurable at Tevatron at low masses to test the model.  
These processes are much smaller than
dijet production (due to the weak QED coupling constant, and to the
small quark component of the Pomeron which initiates these processes),
but they do have appreciable advantages which will appear in the
following discussion. 
The dilepton cross-section is small (same order of magnitude as the diphoton)
but enhanced by the presence of the $Z$ pole (see Fig.~\ref{photonXS}). 
Numerical values 
for the cross-sections are also given in Table \ref{cross}.

\begin{table}
\begin{center}
\begin{tabular}{|c||c|c|c|c|} \hline
Process &(1)&(2)&(3)&(4)\\
\hline\hline
Dijets         & $7.0 10^{5}$ & $3.2 10^{3}$ &$1.2 10^{7}$ &$2.5 10^{6}$ \\
Diquarks         & $3.8 10^{3}$ & $4.4$ & $3.6 10^{4}$ & $1.7 10^{3}$ \\  
$b\bar{b}$     &   $7.6 10^{2}$ &    $8.7 10^{-1}$  &$9.4 10^{3}$ & $4.5 
10^{2}$\\
$\gamma\gamma$ &    $3.5 10^{-2}$ &  $4.9 10^{-5}$  & $4.4 10^{-1}$ & $1.9 
10^{-2}$\\
$l^{+}l^{-}$   &  $4.6 10^{-2}$ & $6.8 10^{-5}$ & $1.1$& $1.3 10^{-2}$\\
\hline
\end{tabular}
\caption{DPE cross-section (pb) for dijets, diquarks, $b \bar{b}$, diphotons 
and dileptons (fb). 
(1): at the Tevatron, $|y|<5$, $m>40$ GeV;
(2): at the Tevatron, $|y|<5$, $m>120$ GeV;
(3): at the LHC, $|y|<5$, $m>40$ GeV;
(4): at the LHC, $|y|<5$, $m>120$ GeV.}
\end{center}
\label{cross}
\end{table}

\subsection{Model predictions for the Tevatron and the LHC: Higgs boson 
production}

\begin{figure}[p]
\begin{center}
\epsfig{figure=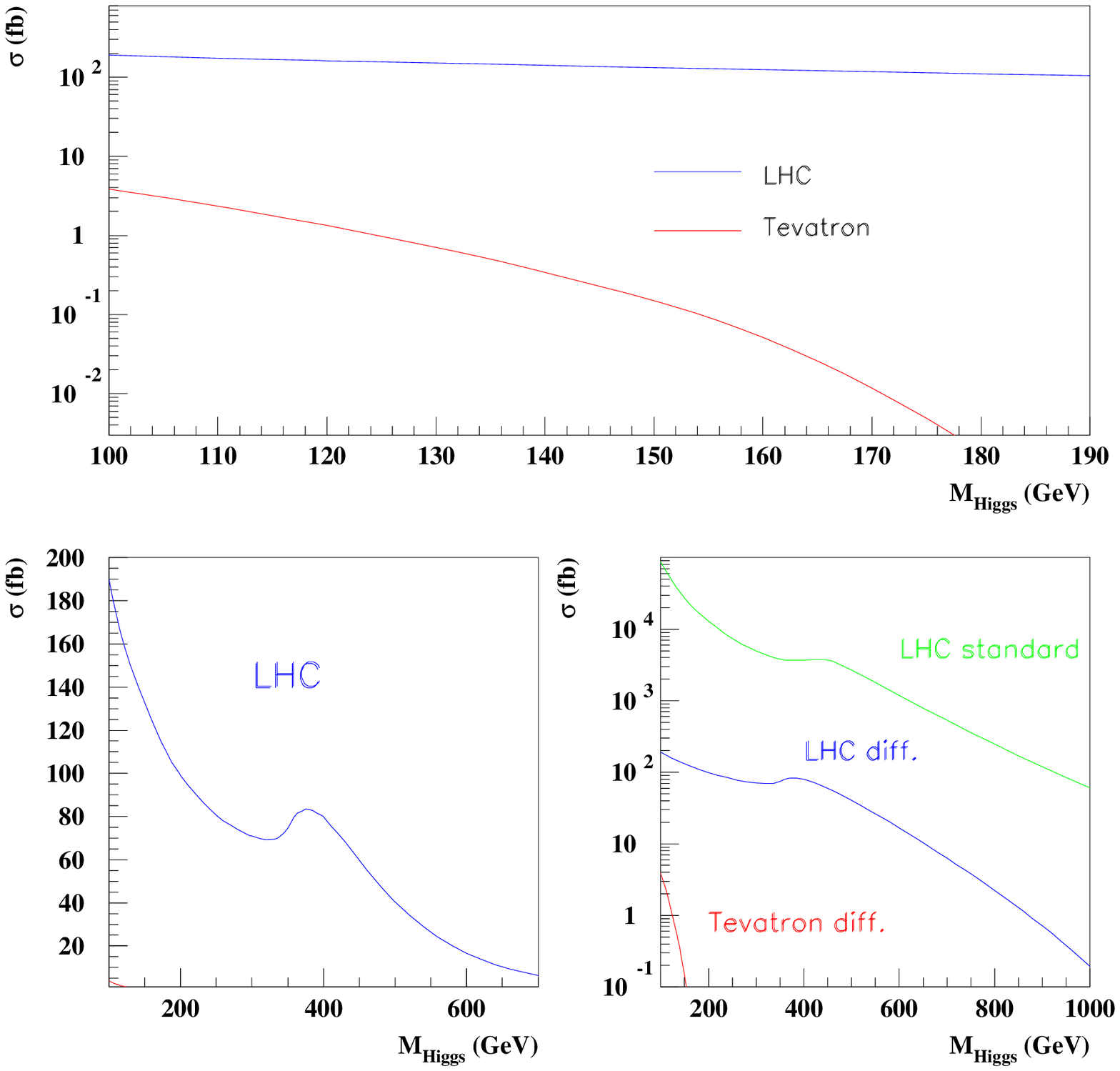,height=6.in}
\end{center}
\caption{Higgs boson production cross-section at the LHC and the Tevatron.
The upper plot gives the cross-sections at the Tevatron and the LHC for low
Higgs mass to show the difference between both accelerators. The bottom plots 
show the same distribution at higher Higgs masses (the standard Higgs 
production
cross-section is shown for comparison).
} 
\label{crosslhcbis}
\end{figure}

\begin{figure}[p]

\begin{center}
\epsfig{figure=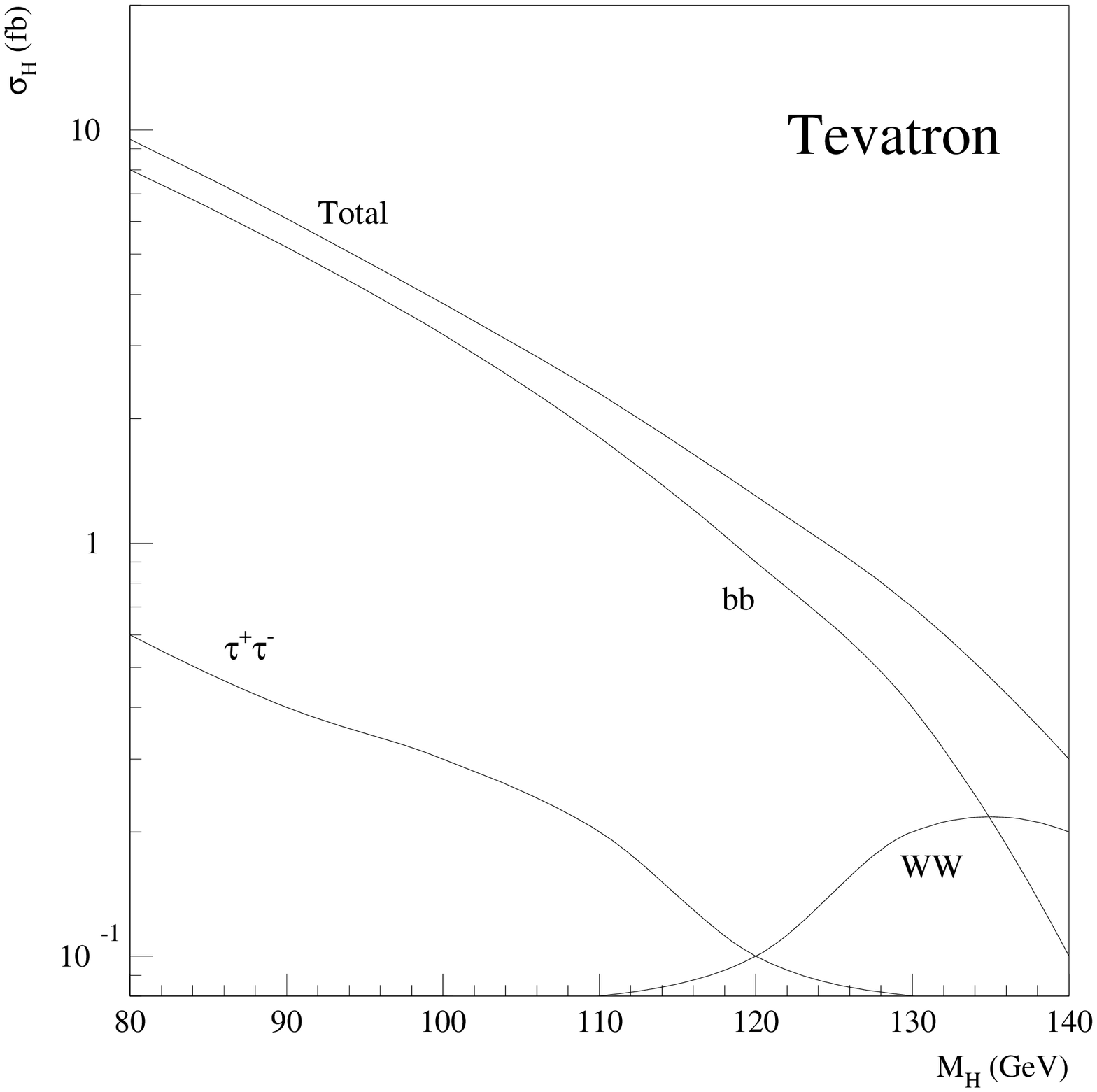,height=3.in} 
\end{center}
\caption{ Higgs boson production cross-section at the Tevatron.
Various decay channels are plotted as a function of the Higgs boson 
mass.}
\label{HiggsXS}

\begin{center}
\epsfig{figure=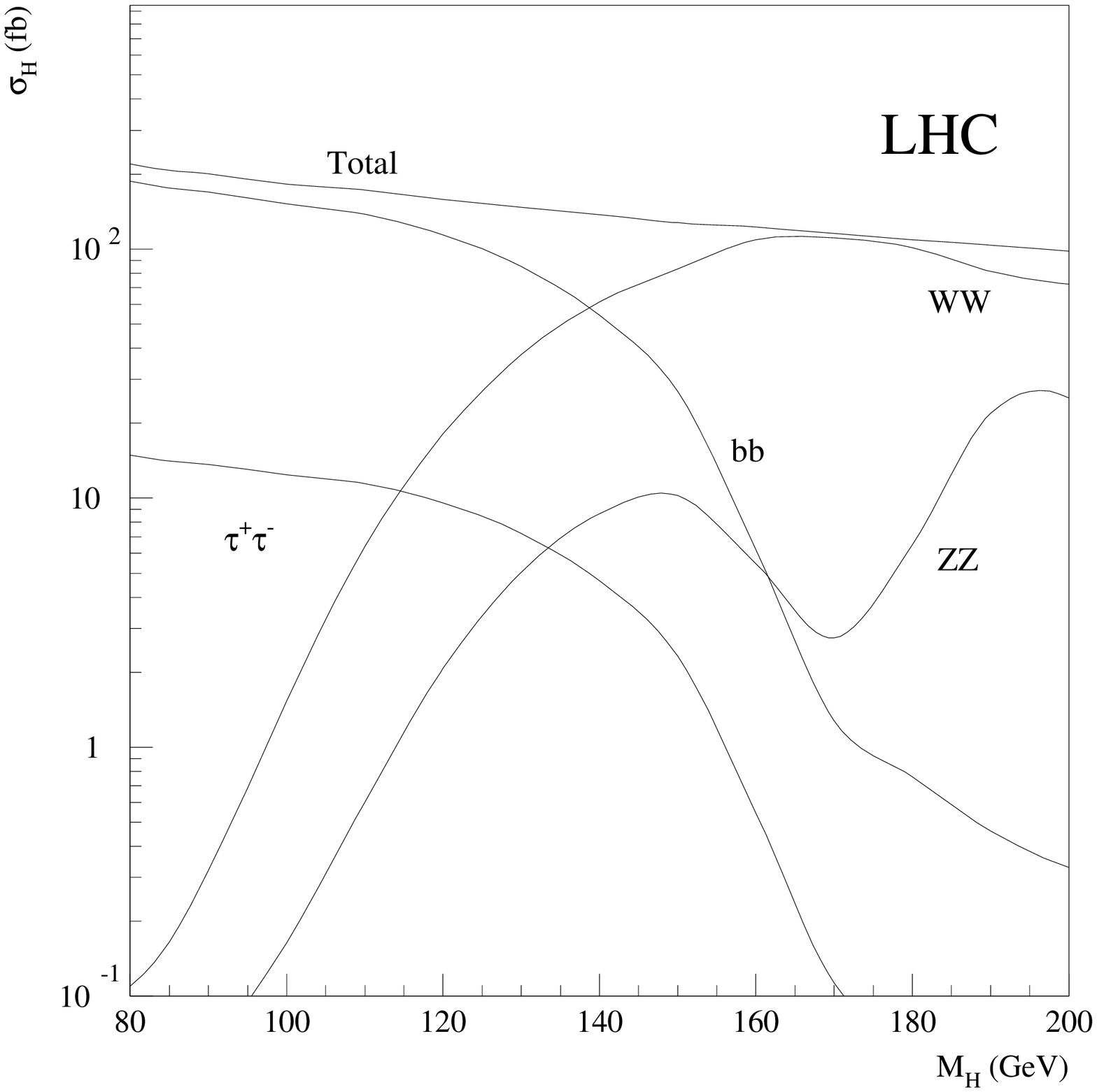,height=3.in}
\end{center}
\caption{ Higgs boson production cross-section at the 
LHC. Various decay channels are plotted as a function of the Higgs boson 
mass.}
\label{HiggsXSb}

\end{figure}

Our results for Higgs boson production at the Tevatron and the LHC are
displayed in Figs.~\ref{crosslhcbis}, \ref{HiggsXS} and
\ref{HiggsXSb}, and in Tables \ref{II} and \ref{III}. 
In Fig.~\ref{crosslhcbis}, we see that the Higgs production cross-section
at the LHC is higher by more than two orders of magnitude than at the Tevatron.
In the same figure is also displayed for comparison the cross-section for
standard Higgs production, which is more than two orders of magnitude larger.
In Figs.~\ref{HiggsXS} and \ref{HiggsXSb}, we give the Higgs boson production 
cross 
section at the Tevatron and the LHC for the different decay modes of the Higgs
boson.

\begin{table}
\begin{center}
\begin{tabular}{|c||c|c|c|c|} \hline
$M_{H}$&(1)&(2)&(3)&(4)\\
\hline\hline
100 & 3.8 & 3.2  & 0.3  & 0.0  \\
110 & 2.3 & 1.8  & 0.2  & 0.0  \\
120 & 1.3 & 0.9  & 0.1  & 0.1  \\
130 & 0.7 & 0.4  & 0.0  & 0.2  \\
140 & 0.3 & 0.1  & 0.0  & 0.2  \\
\hline
\end{tabular}
\caption{DPE Higgs production cross-section at the Tevatron (fb). 
(1): total cross-section, 
(2): $b \bar{b}$ channel,
(3): $\tau\tau$ channel,
(4): $W^+ W^-$ channel.}
\end{center}
\label{II}
\end{table}

The diffractive Higgs production at the LHC is mainly interesting for the lower
Higgs boson masses. When the Higgs boson mass is heavy enough, the $WW$ and 
$ZZ^{(*)}$
decay modes become 
dominant and the visibility of the corresponding channels are already
very good in standard non-diffractive events ; hence we do not expect
the contribution from diffractive channels to be as important there.

At lower Higgs masses, the standard non-diffractive searches are done using the 
$\gamma\gamma$ decay mode, which is loop-mediated and has very small branching 
fraction.
In the present case, the good mass resolution obtained using roman pots or
microstation detectors \cite{orava} allows to use the Higgs boson decays
into b-quarks and into $\tau$-leptons, which are the two main decay
modes when the Higgs boson is lighter than $\sim 140 {\mathrm
GeV/c^{2}}$. We will discuss this point in more detail later in this 
paper.

\begin{table}
\begin{center}
\begin{tabular}{|c||c|c|c|c|} \hline
$M_{H}$&(1)&(2)&(3)&(4)\\
\hline\hline
100 & 182.3 & 152.1 & 12.4 &   1.5 \\
110 & 172.6 & 138.2 & 11.4 &   6.4 \\
120 & 158.5 & 114.3 &  9.6 &  18.1 \\
130 & 147.0 &  85.2 &  7.2 &  37.6 \\
140 & 137.7 &  54.3 &  4.6 &  61.6 \\
150 & 127.5 &  26.8 &  2.3 &  83.4 \\
160 & 122.5 &   6.2 &  0.5 & 109.0 \\
170 & 115.3 &   1.3 &  0.1 & 110.8 \\
180 & 108.9 &   0.8 &  0.1 & 101.4 \\
190 & 103.8 &   0.5 &  0.0 &  81.3 \\
200 &  98.1 &   0.3 &  0.0 &  72.5 \\
\hline
\end{tabular}
\caption{DPE Higgs production cross-section at the LHC (fb). 
(1): total cross-section,
(2): $b \bar{b}$ channel,
(3): $\tau\tau$ channel,
(4): $W^+ W^-$ channel.}
\end{center}
\label{III}
\end{table}

\section{Parameter dependence of DPE cross-sections}
\label{parameter}

In this section, we study the dependence of our predictions for dijet,
diphoton, dilepton and Higgs boson cross-sections on the different parameters 
of the
model. If we refer to the formulae given in the first section, the
relevant parameters of our models are the  momentum transfer slopes
$\lambda_i$ (where $i$ stands for H, JJ, ll, qq, $\gamma \gamma$), and 
$\epsilon$
and $\alpha'$ respectively the intercept and  slope of the Pomeron Regge 
trajectory. We will discuss the dependence on $\epsilon$ in the next
section (note that this value is however imposed in our model, but changes in 
other Pomeron models, as discussed later on).

The dependence on $\lambda_H$, $\lambda_i$
(where $i$ stands for $ll$, $JJ$, $b \bar{b}$, $qq$, and $\gamma \gamma$) is 
displayed in Tables \ref{IV} (for Tevatron) and \ref{V} (for LHC).
In our model, we assumed $\lambda_H=2$ and $\lambda_i=3$ (as in the original 
exclusive model \cite{bi90}). We now vary these
values by one unity, keeping values (assuming or not the equality of the 
$\lambda$'s) around those physically connected to the nucleon form factors.
We also give numbers with the pomeron slope $\alpha'$ taken to be 0.1.
We note as expected that the cross-sections for dilepton, diphoton,
dijet and $b \bar{b}$ production vary by the same factor when we vary the
parameters.

The results of Tables \ref{IV} and \ref{V} are shown for each value of the 
$\lambda$'s 
(except the reference one on the first row) on two lines: on the first line, 
we give the cross-section values
as they come directly from the generator simulation, and on the second lines we 
rescale the values to the reference dijet cross-section\footnote{We remind that 
we 
scale our cross-section predictions to the CDF RunI measurement.}. This method 
allows
us to determine the error on the Higgs production cross-section due to the
assumptions made on the respective values of $\lambda_H$ and $\lambda_i$. For 
the Tevatron,
the cross-section we obtain for a Higgs boson mass of 120 GeV is 1.2 fb, and
varies between 0.6 and 1.9, which gives an incertitude of about 50\% on the
Higgs boson cross-section prediction. For the LHC, the default value is
160 fb and varies between 89 and 220, which gives also an error of about 
50\%.

The other assumption  in our study is to take the gluon density
in the pomeron at $Q^2=75$ GeV$^2$, since this is the upper value of $Q^2$
for the QCD fit. If we take a value at a higher $Q^2$ by using a DGLAP 
evolution
we obtain that the gluon density can vary by a factor 2 at $Q^2=4000$ GeV$^2$.
This has a clear effect on our predictions which can thus vary by a factor 2.

Another assumption of our model which needs to be tested against data is  that
the fraction of events with tagged protons is the same at the LHC and the 
Tevatron, which in 
other words means that the gap survival probability does not depend much
on the center of mass energy. To verify this point will be possible only when 
the LHC energy 
range will be available to experimentation. 

To summarize, we estimate that, on the basis of our model calculations, the 
errors due to the parameter dependence on our predictions for Higgs cross
section (once normalized to dijets) are of the order of a factor 4. 

\begin{table}
\begin{center}
\begin{tabular}{|c||c|c|c|c|c|} \hline
Param. &$\sigma_{dijets}$&$\sigma_{b \bar{b}}$ & $\sigma_{\gamma \gamma}$&
$\sigma_{ll}$& $\sigma_{Higgs}$\\
\hline\hline
$\lambda_H=2$, $\lambda_i=3$ & 7.2 10$^6$ & 2.5 10$^4$ & 1.6  & 1.6
10$^2$ & 1.2  \\ \hline
$\lambda_H=3$, $\lambda_i=4$ & 4.7 10$^6$ & 1.6 10$^4$ & 1.0  & 1.2
10$^2$ & 0.7 \\
 & 7.2 10$^6$ & 2.4 10$^4$ & 1.5  & 1.8
10$^2$ & 1.1 \\ \hline
$\lambda_H=2$, $\lambda_i=2$ & 1.3 10$^7$ & 4.5 10$^4$ & 2.8 & 3.3
10$^2$ & 1.2 \\ 
 & 7.2 10$^6$ & 2.5 10$^4$ & 1.5 & 1.8
10$^2$ & 0.7 \\ \hline
$\lambda_H=4$, $\lambda_i=4$ & 4.7 10$^6$ & 1.6 10$^4$ & 1.0 & 1.2
10$^2$ & 0.42 \\
 & 7.2 10$^6$ & 2.4 10$^4$ & 1.5  & 1.8
10$^2$  & 0.64 \\ \hline
$\lambda_H=1$, $\lambda_i=2$ & 1.3 10$^7$ & 4.5 10$^4$ & 2.8 & 3.3
10$^2$ & 3.5 \\
 & 7.2 10$^6$ & 2.5 10$^4$ & 1.5 & 1.8
10$^2$ & 1.9 \\ \hline
$\alpha'=0.1$ & 9.8 10$^6$ & 3.4 10$^4$ & 2.2 & 2.5
10$^2$ & 1.8 \\
 & 7.2 10$^6$ & 2.5 10$^4$ & 1.6 & 1.8
10$^2$ & 1.3 \\ \hline
\end{tabular}
\caption{ Numerical values of the cross-sections (fb) for different values of 
the parameters,
at the Tevatron 
\newline (the subscript $i$ stands for $ll$, $b \bar{b}$, $JJ$, $\gamma 
\gamma$). 
The Higgs cross-section is for a Higgs mass of 120 GeV.
All other cross-sections are given for a dijet, dilepton, or diphoton mass
greater than 10 GeV.
The first lines read the direct output of the program, whereas the second lines
are rescaled to the same value of the dijet cross-section, namely the CDF
measurement.}
\end{center}
\label{IV}
\end{table}

\begin{table}
\begin{center}
\begin{tabular}{|c||c|c|c|c|c|} \hline
Param. &$\sigma_{dijets}$&$\sigma_{b \bar{b}}$ & $\sigma_{\gamma \gamma}$&
$\sigma_{ll}$& $\sigma_{Higgs}$\\
\hline\hline
$\lambda_H=2$, $\lambda_i=3$ & 4.2 10$^7$ & 1.1 10$^5$ & 8.8 10$^{3}$ & 1.6
10$^3$ & 1.6 10$^2$  \\ \hline
$\lambda_H=3$, $\lambda_i=4$ & 2.7 10$^7$ & 7.5 10$^4$ & 5.5 10$^{3}$ & 1.0
10$^3$ & 9.0 10$^1$  \\ 
 & 4.2 10$^7$ & 1.2 10$^5$ & 8.6 10$^{3}$ & 1.6
10$^3$ & 1.4 10$^2$  \\ \hline
$\lambda_H=2$, $\lambda_i=2$ & 7.1 10$^7$ & 1.9 10$^5$ & 1.6 10$^{4}$ & 2.8
10$^3$ & 1.6 10$^2$  \\ 
 & 4.2 10$^7$ & 1.1 10$^5$ & 9.5 10$^{3}$ & 1.7
10$^3$ & 9.5 10$^1$  \\ \hline
$\lambda_H=4$, $\lambda_i=4$ & 2.7 10$^7$ & 7.5 10$^4$ & 5.5 10$^{3}$ & 1.0
10$^3$ & 5.7 10$^1$  \\ 
 & 4.2 10$^7$ & 1.2 10$^5$ & 8.6 10$^{3}$ & 1.6
10$^3$  & 8.9 10$^1$  \\ \hline
$\lambda_H=1$, $\lambda_i=2$ & 7.1 10$^7$ & 1.9 10$^5$ & 1.6 10$^{4}$ & 2.8
10$^3$ & 3.8 10$^2$  \\ 
& 4.2 10$^7$ & 1.1 10$^5$ & 9.5 10$^{3}$ & 1.7
10$^3$ & 2.2 10$^2$  \\ \hline
$\alpha'=0.1$ & 6.1 10$^7$ & 1.7 10$^5$ & 1.5 10$^{4}$ & 2.3
10$^3$ & 2.5 10$^2$  \\ 
 & 4.2 10$^7$ & 1.2 10$^5$ & 1.0 10$^{4}$ & 1.6
10$^3$ & 1.7 10$^2$  \\ \hline
\end{tabular}
\caption{Numerical values of the cross-sections (fb) for different values of the 
parameters
at the LHC ($i$ means $ll$, $b \bar{b}$, $JJ$, $\gamma \gamma$). 
The Higgs cross-section is for a Higgs mass of 120 GeV.
All other cross-sections are given for a dijet, dilepton, or diphoton mass
greater than 10 GeV.
The first lines read the direct output of the program, whereas the second lines
are rescaled to the same value of the dijet cross-section, namely the CDF
measurement.}
\end{center}
\label{V}
\end{table}

\section{Testing and comparing  models at the Tevatron Run II}
\label{sectV}

Let us discuss and propose ways to verify our model and 
determine its parameters more precisely using the Tevatron run II 
data for dijets, diphotons and dileptons. This will allow to make improved 
predictions for 
{\it e.g.} Higgs boson production later on at the LHC. In the same time, this 
will give  the possibility to 
differentiate, and thus compare the existing models which give predictions for
diffractive Higgs production. 

\subsection{Brief discussion of other models}

Models attempting to describe central diffractive production from $p \bar{p}$ or 
$pp$
interactions  base their predictions on
either an explicit color singlet exchange of two gluons (where one may be
soft, as in our model), or in terms of 
hadronic  interactions, in which diffractive features (rapidity gaps,
leading protons)  appear in relation with soft initial and/or final state 
interactions.

In this section we attempt to summarize results obtained in 
different pictures for the same observables, and assess how observation could 
allow to
distinguish them. We concentrate on recent models\footnote{In fact, the 
observable tests that we propose could be applied to other models as well, and 
thus have a wide applicability.} which have provided
explicit numbers for various processes of interest (namely dijet,
diphoton and Higgs boson production) so that a comparison of predictions is made
possible. We therefore compare our results with the {\it exclusive} model
predictions of Ref. \cite{allinclu}-a, with the double Pomeron
{\it inclusive} model of Refs. \cite{allinclu}-b
and with the soft color interaction models of Ref. \cite{allinclu}-c. These 
three 
approaches will be shortly described below.

Two-gluon models for exclusive production have been first proposed, ({\it cf.}, 
see \cite{bjorken}), but  led in general to too large cross-sections, in
conflict with the upper bound from CDF dijets. Despite their initiatory r\^ole 
in the problem, they will not be discussed in
detail here. We cannot either include the
original Bialas-Landshoff exclusive model \cite{bi90} in this discussion, since
the model has not been extended to gluon pair production. A
confrontation of the exclusive Bialas-Landshoff dijet production with
the CDF upper bound is thus not possible at this point.

Exclusive production is evaluated in \cite{allinclu}-a as a 
perturbative process involving two-gluon exchange tamed by Sudakov suppression 
factors supplemented by rapidity-gap suppression. Two gluons are extracted from 
the beam particles
according to the usual proton structure functions (the process can thus be 
interpreted as
``proton-induced'') and one of them couples to the hard central system
(e.g. a high mass dijet, Higgs boson {\it etc}...). The infrared divergence of
this process ($d\sigma \propto dQ_{t}^{2}/Q_{t}^{4}$, $Q_{t}$ being the gluon
transverse momentum) is regularized by the requirement that no radiation
occurs along the gluon which would fill the rapidity gap. This  leads to 
Sudakov-like form factors  effectively damping the low-$Q_{t}$ region, so that 
the
whole process can be considered as  entirely perturbative and thus calculable. 
Finally a conventional rapidity gap survival factor corrects for the soft 
rescattering corrections. Clues for future
studies on distinguishing the exclusive from the inclusive DPE
processes are given in Section \ref{exc-inc}.

The model for inclusive DPE presented in \cite{allinclu}-b, and implemented in
{\tt POMWIG} \cite{cox} is a direct transposition of the Pomeron flux and
parton densities from the $ep$ to the $pp$ context, and is therefore
referred to as a factorizable model\footnote{As in our model, the model is
Pomeron-induced. However, our model is based on a soft Pomeron 
interaction between the initial hadrons which do not verify Regge factorization 
at the incident vertices. Here the model is based on a 
Regge-factorized mechanism with a Pomeron at each vertex, the necessary 
factorization breaking coming from the gap suppression factor.}. In particular, 
it uses 
the full Regge parametrization of Pomeron flux factors as determined
on the HERA data quoted before. The factorizable model is formally
similar to ours, since both models satisfy QCD factorization by the use of 
gluon 
structure functions in the Pomeron to describe the hard process in the central 
rapidity region. However, the choice of parameters and color structure  
reflects 
a different interpretation of the nature of the
Pomeron, as outlined in section \ref{inc-inc}. The closely related formulation
of both  models will allow to discuss their differences in terms of
physical parameters.

Two models have been designed to describe diffractive physics as soft
color rescattering over a hard subprocess \cite{allinclu}-c, namely SCI (Soft 
Color Interaction)
and GAL (General Area Law). They are implemented as a transition
between the hard interaction and hadronization, and therefore fit
naturally in Monte-Carlo programs such as {\tt PYTHIA} \cite{PYTHIA}. 
In these models, color exchanges occur in the final state, potentially
stopping color flow between the remnants of the incoming hadrons and
the central system, leading to diffractive event topologies. These final
state interaction models have a different formulation from the
previous ones\footnote{Note a formal similarity with the 
Bialas-Landshoff mechanism with the superposition of  hard and  soft color 
exchanges in the same probability amplitude.}. We will restrict ourselves to 
recall their predictions
for various DPE final states, insisting on differences with the other models.

\subsection{Exclusive ``proton-induced'' model}
\label{exc-inc}

The question of the observation of exclusive DPE events is open. As we
have said, the current data allow only to place an upper bound on the
process, and this bound is still compatible with (but not far lower from) the 
predictions of
\cite{allinclu}-a. The forthcoming Tevatron Run-II data will give new insight on
this question, by looking for diffractively produced central dijets,
or even better diphotons and dileptons since the events will be cleaner. 

\begin{figure}[p]

\begin{center}
\epsfig{figure=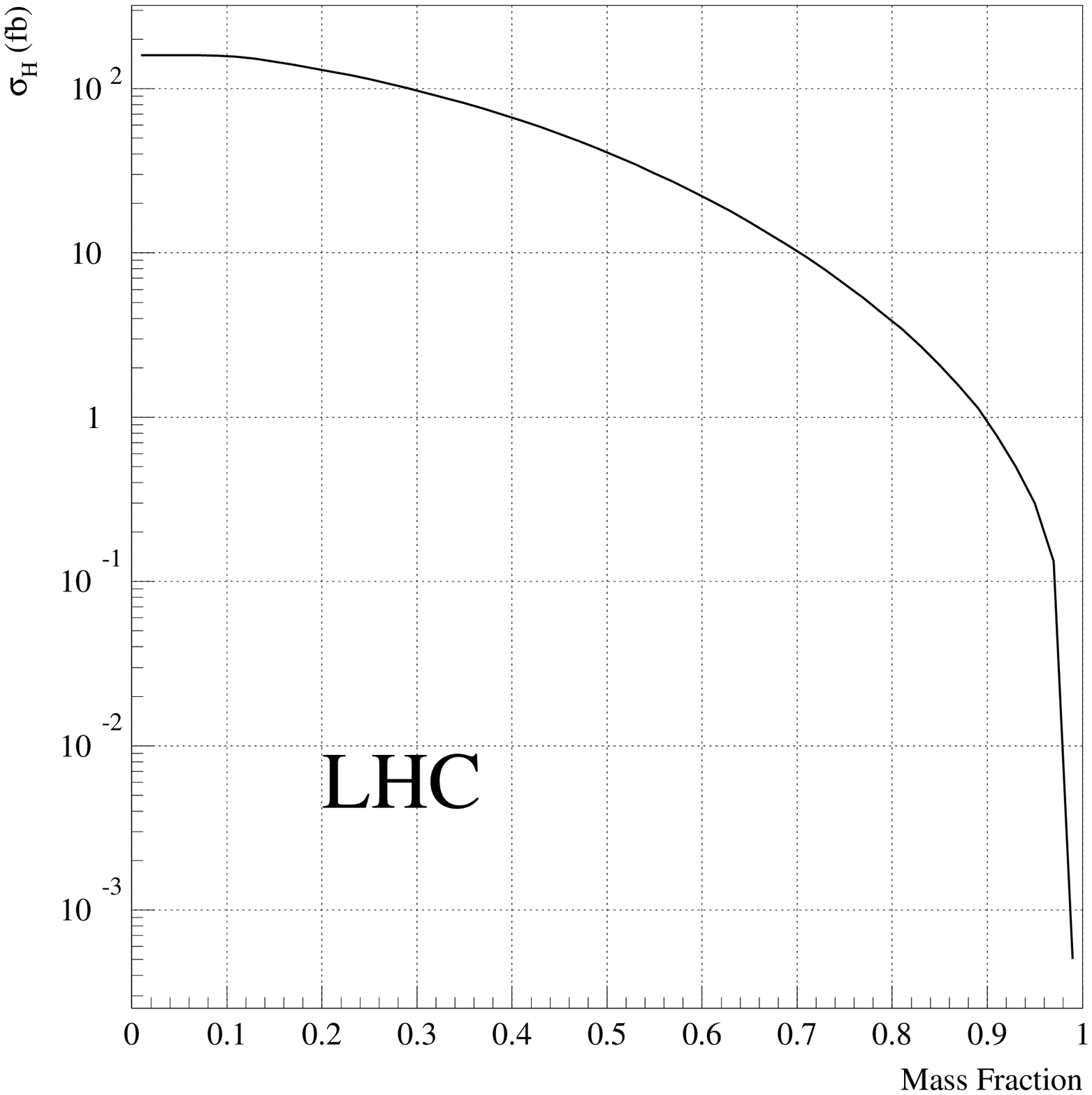,height=3in}
\end{center}
\caption{Remaining diffractive Higgs boson cross-section at the LHC, as a
function of the cut on the mass fraction.}
\label{massfrac}

\begin{center}
\epsfig{figure=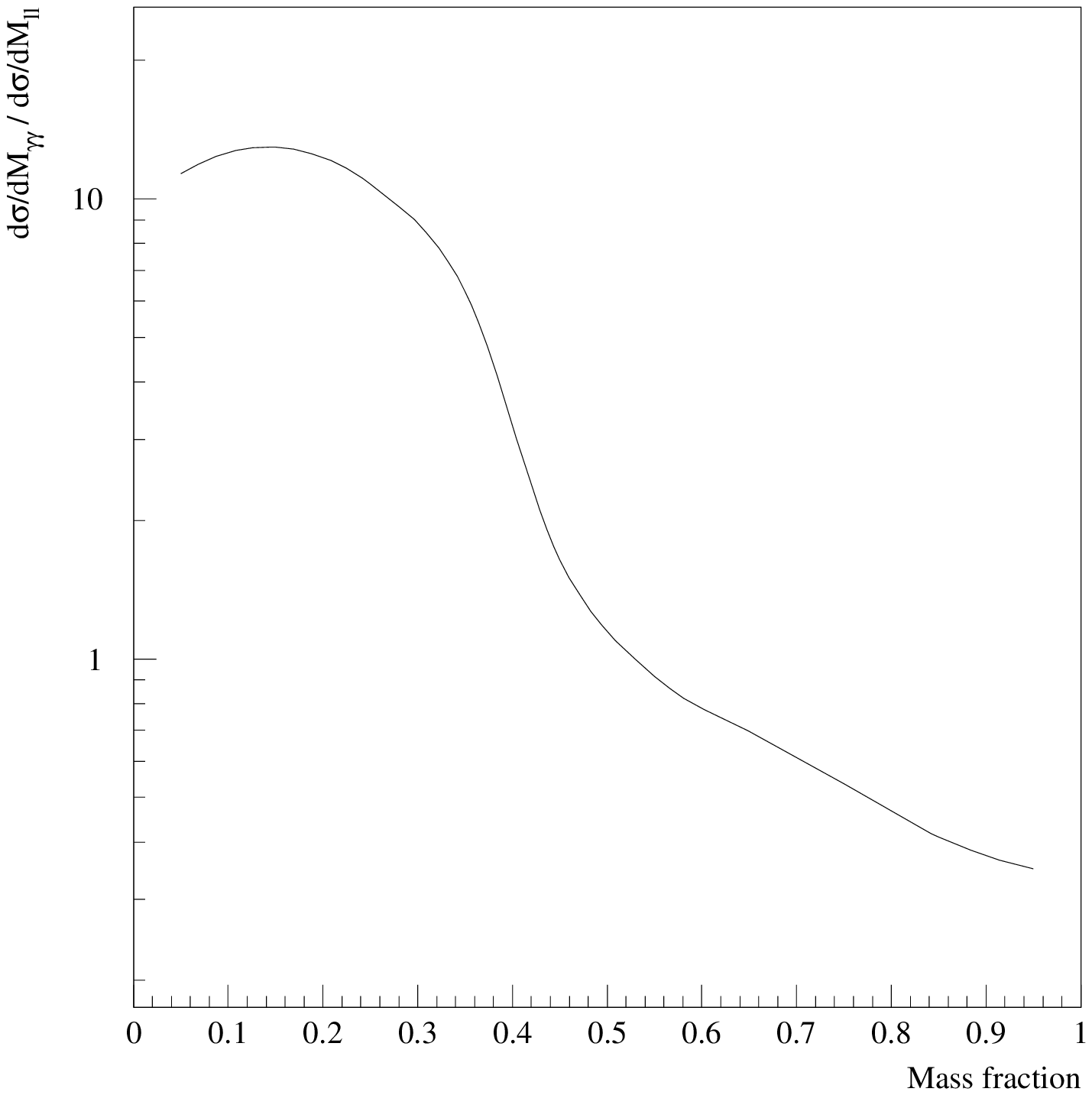,height=3in}
\end{center}
\caption{Diphoton to dilepton cross-section ratio, as a function of the mass 
fraction.}
\label{ggllratio}

\end{figure}

A specific question concerns the separability of the exclusive process from
the inclusive background. The exclusive processes will show a high value
of the dijet, dilepton or diphoton mass fraction. In our model, these
events correspond to the case when the pomeron remnants show very little
energy. It is thus worthwhile to verify whether one is able to distinguish 
between these 
two configurations. If we assume an experimental resolution of $15\%$ on the 
measured dijet mass, one should require $M_{frac} > 0.85$ for both the exclusive 
or
the inclusive events. With this cut, we obtain a cross-section from our 
inclusive
model of about 2 fb at LHC, which is large enough to be
competitive with the exclusive numbers given in \cite{allinclu}-a. The Higgs 
boson cross-section 
as a function of the mass fraction is given in Fig.~\ref{massfrac}. In fact, 
this 
shows that it will be difficult to distinguish between pure exclusive events and
the tail of the inclusive events which we can call ``quasi-exclusive'' events
\footnote{Let us mention another source of backgrounds to the exclusive 
process,
coming from the experimental difficulty of telling that a tagged
forward proton is not resulting from an $N^{*}$ decay and accompanied
by other undetected particles. Provided the decay proton is still in
the required momentum range ($\xi < 0.1$), this will result in an
underestimation of the missing mass, $i.e.$ an overestimation of the
mass fraction, hence polluting the selection.}.
It is thus important to be able to tag both kinds of events \cite{usbis}.

Another example of discriminating observable is given by the ratio of the
diphoton and dilepton cross-sections, as a function of the mass-fraction
value. In the inclusive models, this ratio is determined by the quark
and gluon distributions inside the Pomeron, and the presence of the Z
pole in the dilepton cross-section. It is illustrated in Fig.~\ref{ggllratio}. 
The presence of an exclusive contribution to the
cross-section would result in a sharp enhacement at $M_{frac} \sim 1$, since
diphotons are allowed in the exclusive case, whereas dileptons are absent.

In any case, measurements of the cross-section of exclusive or
quasi-exclusive events at Tevatron, in the case of dilepton, diphoton or
dijet production will provide a direct test of the models.  

\subsection{Comparison  of the ``Pomeron-induced'' models  at the Tevatron Run 
II}
\label{inc-inc}

It has already been noticed that our model presented here, and the factorizable
Pomeron induced model of \cite{allinclu}-b in practice differ in the relative 
normalizations of Higgs boson {\it vs.} dijets and
in the value of the parameters. In particular, we take $\epsilon=0.08,$ see 
formula 
(\ref{dinclujj})-(\ref{dinclugg}) while the factorizable model
assumes a value of the pomeron intercept $\epsilon=0.2$ coming from
the HERA measurements. Also,  the slope in
momentum transfer originates  from the nucleon form factor in our model 
(eventually supplemented by an additional term related to the
non-perturbative gluon propagator)
whereas the factorizable
Pomeron induced model takes the value from HERA measurements. It is 
thus interesting to vary this parameter.

Note an important difference between model predictions due to the distinctive  
color reconnection pattern 
of the two models. In the factorized model the Higgs boson to dijet
normalization ratio is: $
\left(\frac{C_{H}}{C_{JJ}}\right)_{fact} = \frac{\sqrt{2}G_{F}}{24\pi},$ with a 
factor 8 smaller than (\ref{ratio}) for the 
non-factorized model. The reason is that the factorizable model authorizes 
dijets  coming from two quarks or gluons bearing different color charges, while 
they are restricted to be in an overall singlet state in the non-factorizable 
model due to the common gluon  exchanged between the incident 
nucleons (in other terms the whole of Pomeron remnants is essentially color 
singlet). Hence the prediction of the Higgs boson {\it vs.} dijet cross-section 
ratio keeps track of this difference. Here again, insight can be gained from the 
comparison 
of diphoton (or dilepton) and dijet cross-sections: since the first ones have 
the same
color structure as Higgs boson production, they can be used to infer which, of 
the 
factorizable and non-factorizable pictures, is correct.

\begin{figure}[p]

\begin{center}
\epsfig{figure=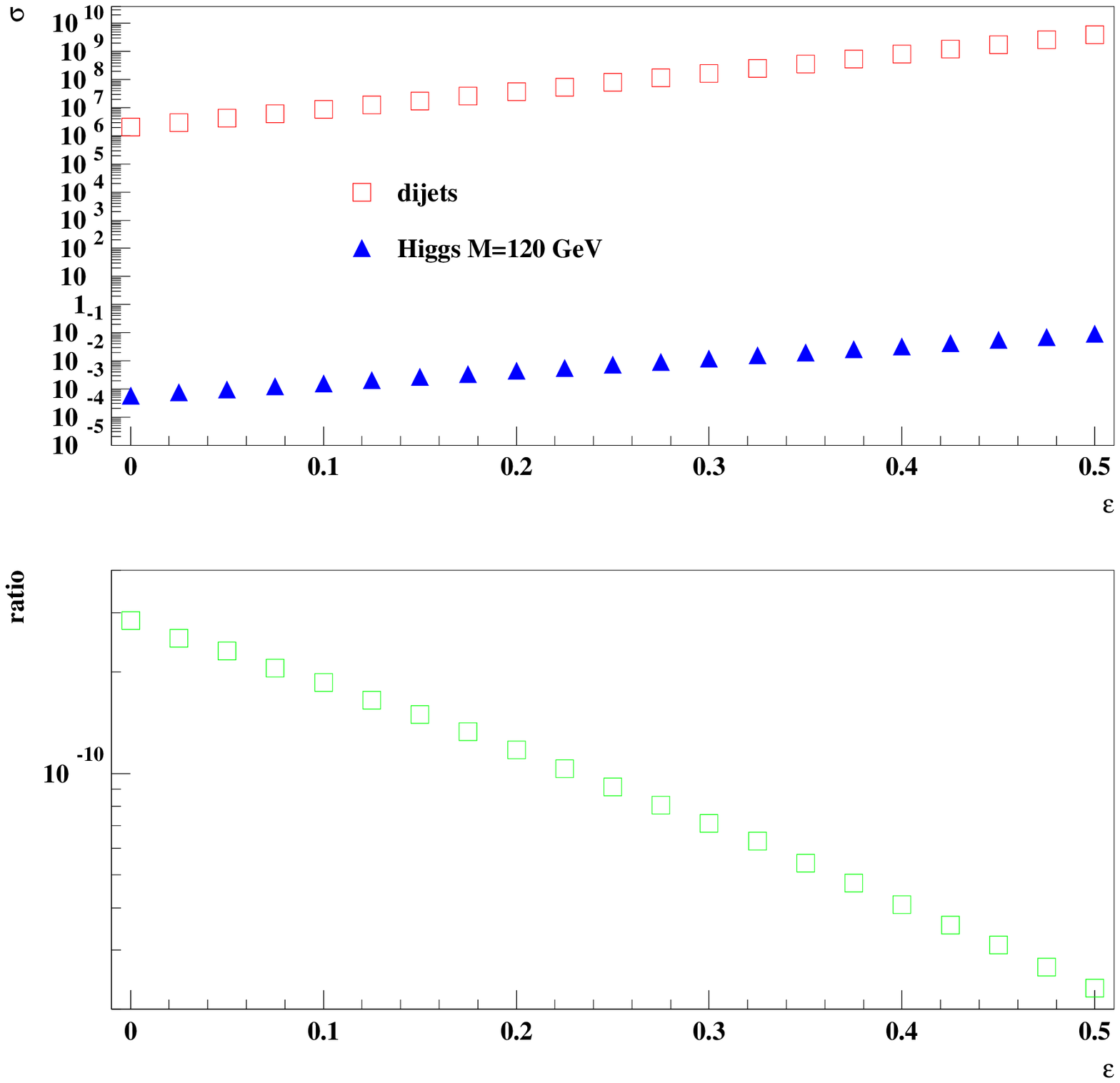,height=3.in}
\end{center}
\caption{Higgs boson and  dijet production at Tevatron: raw  cross-sections and 
their ratio for different 
values of $\epsilon$, at $M_H=120 {\rm GeV}$.}
\label{epsilon1}

\begin{center}
\epsfig{figure=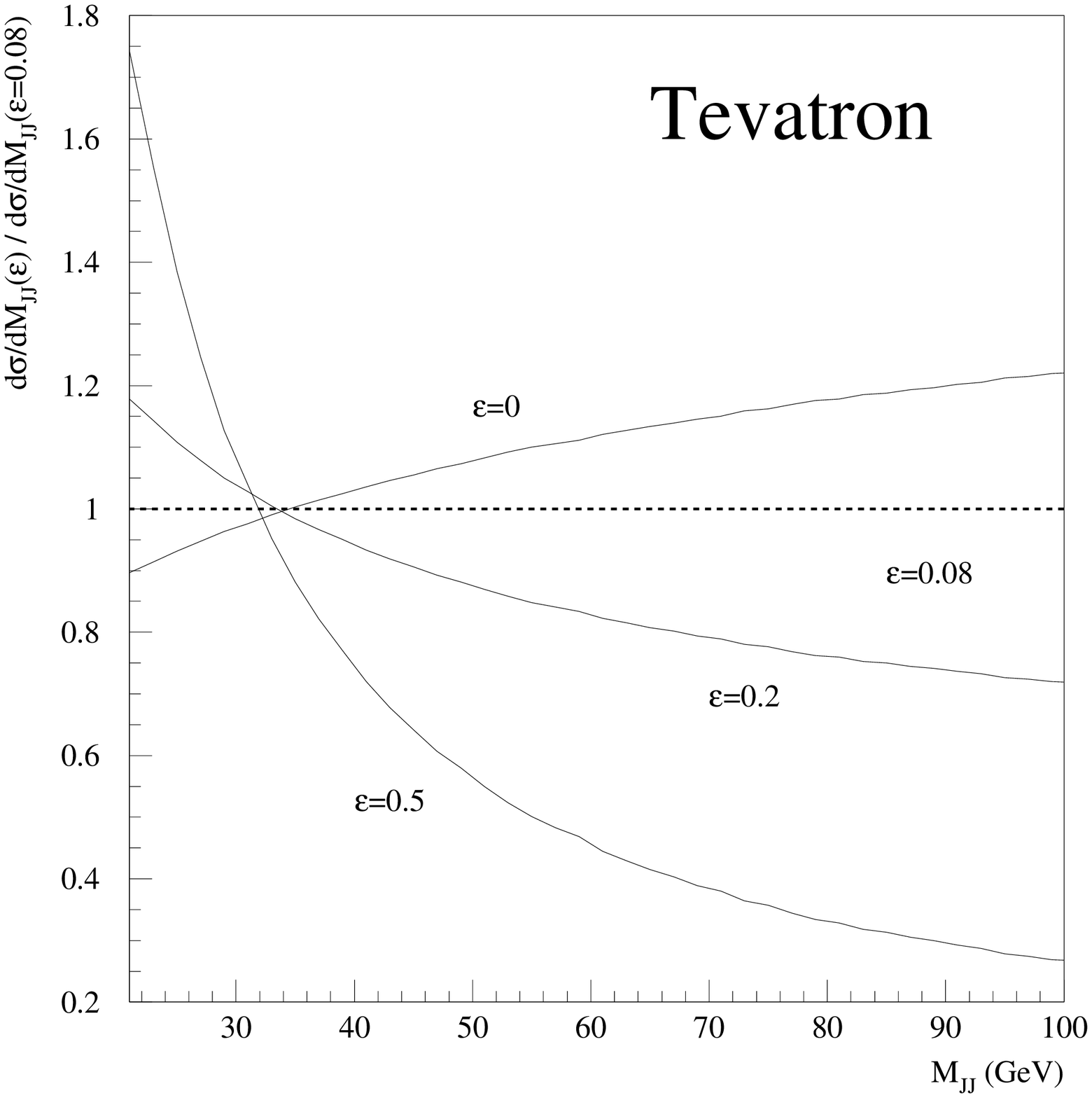,height=3.in}
\end{center}
\caption{Differential dijet cross-sections for different 
values of $\epsilon$, compared to a reference taken at $\epsilon= 0.08$. All 
cross-sections are normalised to the same value. The values $\epsilon= 0.08$, 
0.2 and 0.5 correspond,  
respectively, to the soft, hard (measurement at HERA), and hard BFKL
pomeron intercepts.}
\label{epsilon2}

\end{figure}

Fig.~\ref{epsilon1} shows the variation of the Higgs and dijet
cross-sections as a function of $\epsilon$ at the generator level,
and their ratio. As can be guessed from
formulae (\ref{dinclujj})-(\ref{dinclugg}), an increase of $\epsilon$ will 
enhance the
cross-section essentially at low diffractive mass, therefore the effect
on the dijet cross-section is much larger than on the Higgs boson
cross-section. When $\epsilon$ is increased, the dijet cross-section is
increased, and thus the multiplicative factor needed to adjust the 
prediction to the CDF Run I measurement is smaller. This explains why
the predicted Higgs cross-section is expected to decrease when 
$\epsilon$ increases.

\begin{figure}[p]

\begin{center}
\epsfig{figure=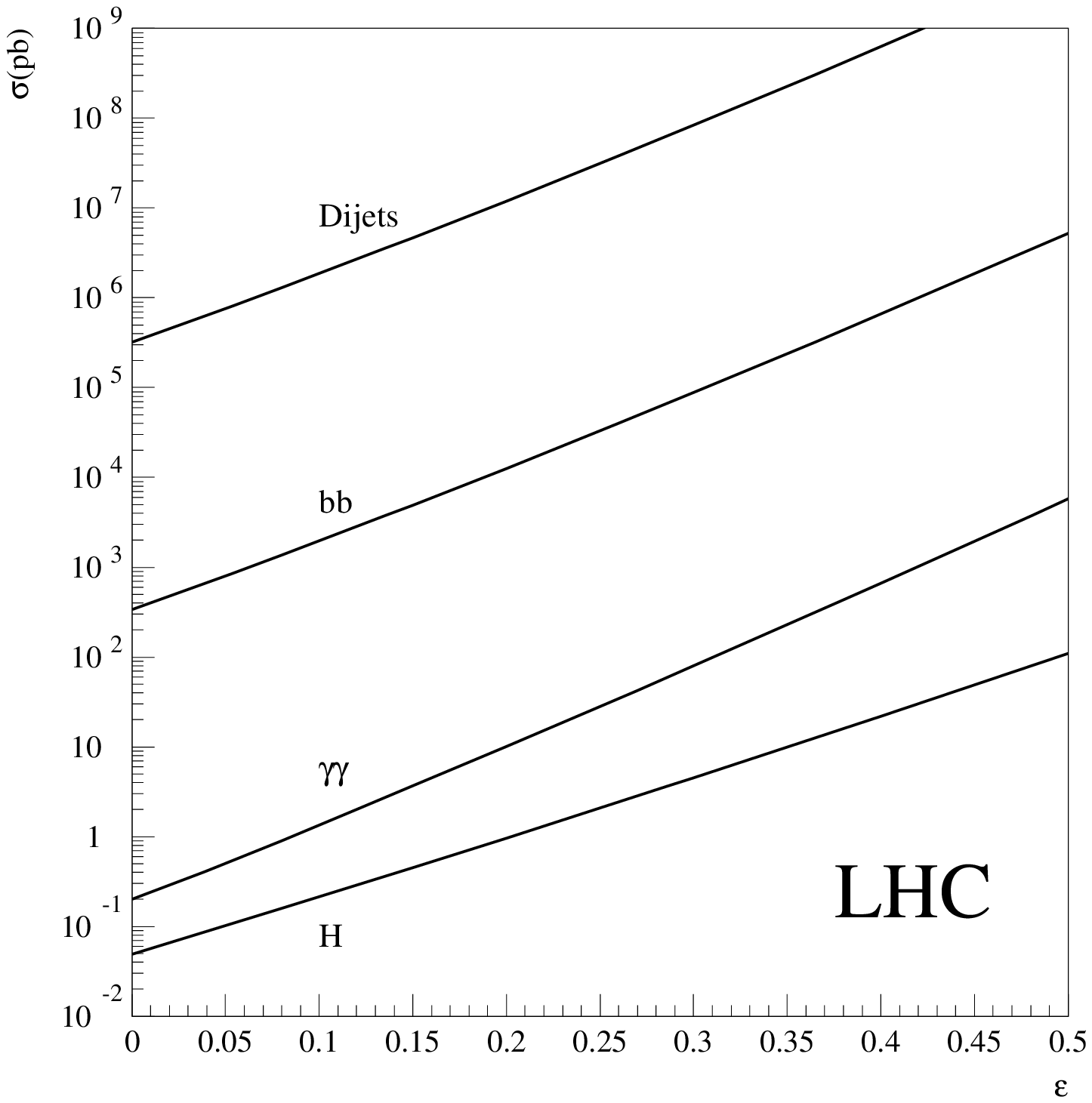,height=3in}
\end{center}
\caption{Raw cross-sections vs. $\epsilon$}
\label{epsilona}

\begin{center}
\epsfig{figure=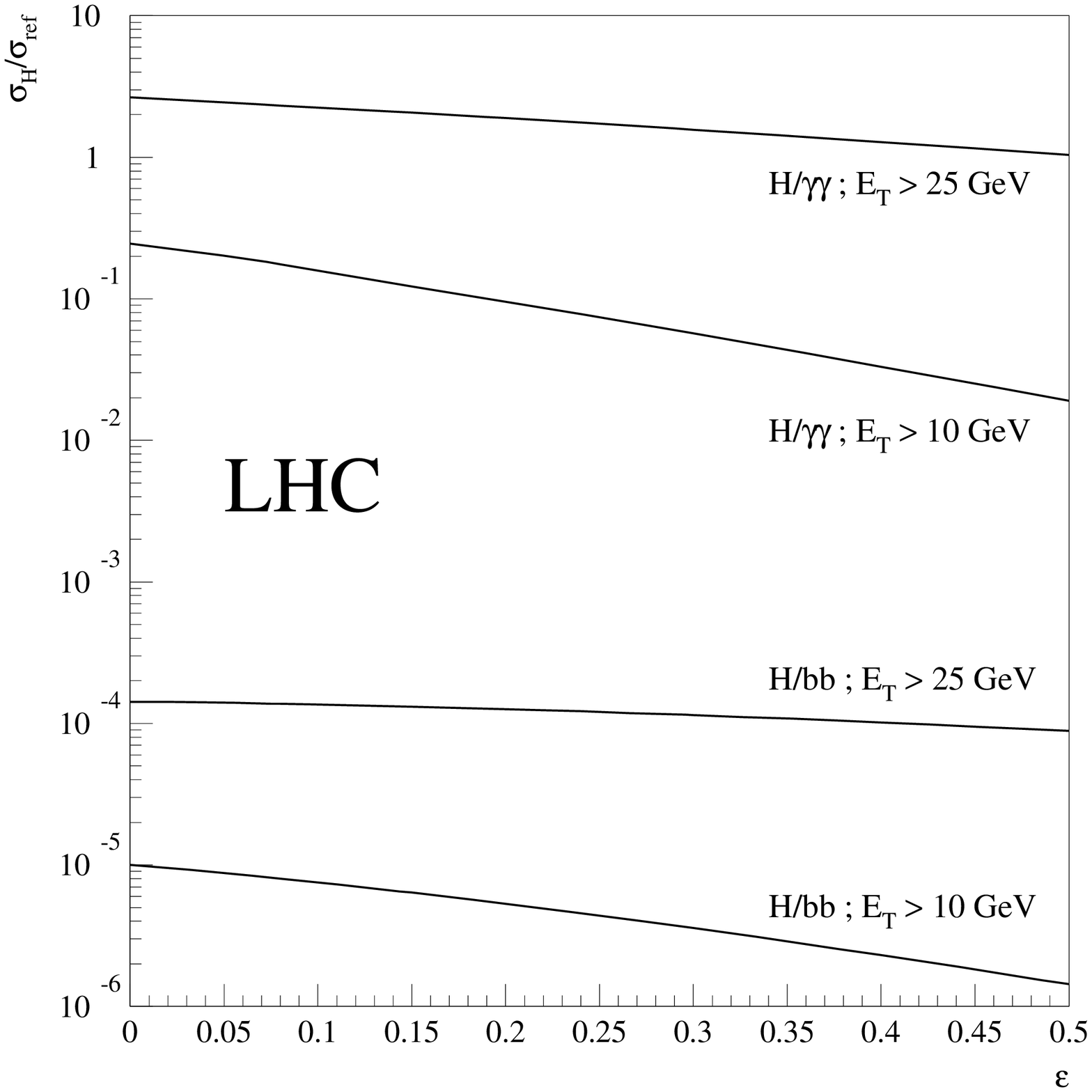,height=3in}
\end{center}
\caption{Higgs compared to dijets and diphotons as a function of $\epsilon$}
\label{epsilonb}

\end{figure}

Fig.~\ref{epsilon2} shows the $\epsilon$-dependence of the differential dijet
cross-section $d\sigma_{JJ}/dM_{JJ}$, when compared to ($i.e.$, divided by) a 
reference 
distribution, taken at $\epsilon= 0.08$. A strong dependence is found,
meaning that the study of this distribution, independently from its 
normalisation, should 
allow to constrain the value of $\epsilon$, and to obtain a more precise 
prediction 
for the Higgs boson cross-section at the 
LHC\footnote{The sensitivity of $d\sigma_{JJ}/dM_{JJ}$ to
$\epsilon$ will however crucially depend on the experimental determination
of the jet energy scale and on the remaining uncertainties on the
Pomeron structure function $G_{P}$.}
Similar studies can be done with dilepton and 
diphoton events, which will benefit from much central mass resolution.

In Fig.~\ref{epsilona} we give the cross-sections for dijets, $b \bar{b}$,
$\gamma \gamma$, and Higgs boson, and in Fig.~\ref{epsilonb}, the ratio of the 
Higgs
cross-section with respect to $\gamma \gamma$ and $b \bar{b}$ as a function of
$\epsilon$ for the LHC. Namely, it will allow to get a prediction on these
cross-sections, once   $\epsilon$ been determined, {\it e.g.} at the Tevatron.

These last remarks highlight once more the interest of DPE diphoton and
dilepton production. Indeed, the experimental resolution on photon,
electron and muon momenta is much better than the jet energy resolution, 
since in this case the events benefit from tracking information and/or
electromagnetic calorimetry with a much better defined energy scale.
Moreover, especially in the dilepton case, the hard sub-process is
essentially quark-initiated (see formula (\ref{dinclull})), and the knowledge of 
$Q_P$ is much more
precise than that of $G_{P}$, thus eliminating another source of uncertainty.
The measurement of $\epsilon$ will also be possible using the diphoton
and dilepton measurement and the dependence of the cross-sections as a
function the diphoton or the dilepton mass.

\subsection{Predictions: summary and comparison}

In Table \ref{comp}, we give, where available, the cross-sections obtained by 
the different
models. The factorizable numbers at the LHC (second column) should contain a 
factor accounting 
for the unknown rapidity gap survival probability (denoted $SP$); this factor is 
around 0.1 
at the Tevatron. The numbers from the proton-induced model refer to $exclusive$ 
production, and 
are as such not refering to strictly the same process as the pomeron induced 
ones. 
The various models have fundamentally different philosophies and 
assumptions, so that caution should be taken not to conclude from any numerical 
agreement that 
the models themselves agree.

\begin{table}
\begin{center}
\begin{tabular}{|c||c|c|c|c|} \hline
process and cuts &(1)&(2)&(3)&(4)\\
\hline\hline
Higgs, 115 GeV, Tev & 1.7 & 0.029-0.092 & 0.03 & 0.00012 \\
Higgs, 115 GeV, LHC & 169. & 379.-486. $\times SP$ & 1.4 & 0.19 \\
Higgs, 160 GeV, LHC & 123. & 145. $\times SP$ & 0.55 & - \\
$\gamma \gamma$, Tev, $E_T>12 GeV$, $\eta <2$ & 71. & 128. (27.) & - & $\sim$ 
20.\\
$\gamma \gamma$, Tev, $E_T>12 GeV$, $\eta <1$ & 9. & 8. (2.) & - & - \\
$\gamma \gamma$, LHC, $E_T>50 GeV$, $\eta <2$ & 1.5 & - & - & 0.1 \\
$\gamma \gamma$, LHC, $E_T>120 GeV$, $\eta <5$ & 19. & - & 0.12 & -\\
\hline
\end{tabular}
\caption{DPE Higgs and diphoton production cross-section at the LHC and
the Tevatron (fb) for different models.  
(1): non factorisable pomeron based model (present work), 
(2): factorisable pomeron based model (the results contain
both the pomeron and the reggeon components. For
the $\gamma \gamma$ cross-section, we give in parenthesis the contribution
of the pomeron only). 
(3): proton based model (exclusive case; the results are given for a Higgs
mass of 120 GeV),
(4): soft color interaction models.}
\end{center}
\label{comp}
\end{table}

\section{Advantage  of diffractive {\it vs.} ``standard'' Higgs boson 
production}
\label{sectVI}

As is now well-known \cite{bi90,bjorken,albrow} \cite{allinclu}-a, exclusive DPE 
events
are very attractive
from the experimental point of view, since the measurement of the
outgoing protons momenta, using dedicated forward detectors (like roman
pots or microchambers \cite{orava}) allows to determine the mass of the
centrally produced system with excellent precision, so that the Higgs
boson signal can be extracted from the DPE continuum
background. Moreover, the background itself is strongly suppressed due 
to angular momentum conservation constraints (the ``$J_Z=0$'' rule
\cite{pumplin}, see section \ref{formulation}). Exclusive DPE events
are however not an experimentally established process according to the
data currently at our disposal. As we have emphazised, the
``quasi-exclusive processes'' which are quite close
in topology from the exclusive processes are established processes and may keep 
an essential part of the nice kinematical features of the exclusive production.

In general inclusive DPE events, the angular momentum constraint above does not
apply and the background is not suppressed, so that the {\it a priori}
signal-to-background ratio ($S/B$) is similar to that of ``standard'' 
non-diffractive 
events. The missing mass method will not work as such, since the
equality between the missing mass to the outgoing protons and the
mass of the central system does not hold.

However, inclusive DPE events can still be kinematically constrained in
a much stronger way than non-diffractive events. If, in addition to the
forward proton detectors, the experiments can dispose of very forward
calorimetry (e.g. up to $\eta\sim 8$, as was proposed in \cite{forwardcalo}),
one may use the measurement of the Pomeron remnants to fully constrain
the events kinematics. Combined with the higher cross-sections predicted 
in this last case, it shows that such forward calorimetry would be
adapted and necessary. The first resolutions using these detectors
are given in \cite{usbis}.

The idea is to use the forward calorimeters as vetos. The
central mass is reconstructed with the proton momenta only, and events
are considered only if the measured remnant energy is small enough. In
a sense, this is a way to select ``quasi-exclusive'' events. Fig.~3
of Ref. \cite{usbis} shows the achievable resolution as a function of the 
remnant
veto.

Let us finish this discussion with an important point concerning the
interest of the $\tau\tau$ decay mode. According to section
\ref{modpred}, the inclusive DPE $\tau\tau$ cross-section is very
small compared to the DPE dijet cross-section, and this is understood
from the smallness of the QED coupling constant and the quark component
of the Pomeron. Compared to this, the Higgs boson branching fraction
into $\tau$ pairs is still $\sim 10\%$ in the mass range we
consider. It follows from the cross-sections given earlier (see Table \ref{III}, 
and 
Fig.~\ref{HiggsXSb})
that the cross-section ratio for $\tau$-pair events is ${\cal O}(10)$ at for 
example
$m_{\tau\tau} \sim 130$ \,GeV for inclusive DPE events, whereas this
mode is invisible through non-diffractive events.

Of course, the missing energy in $\tau$ decays, the presence 
of the Z peak in the $\tau$-pair cross-section, and other background processes 
not considered here spoil these numbers to
some extent, but we do consider that this channel needs investigation. Also
note that a pseudoscalar Higgs boson A, which appears in models with an
extended Higgs sector (such as the supersymmetric models), does not
couple to  gauge bosons and hence preserves a significant branching
fraction into $\tau$-pairs up to very high masses, where the effect of
the Z peak is negligible; the same remark is true for the MSSM h boson in some 
regions of the
parameter space.

\section{Summary}
\label{sectVII}

The present work is meant to serve as a reference to the non factorizable, 
Pomeron induced DPE model. 
\begin{itemize}
\item The original exclusive DPE formulation are  recalled, and the inclusive 
expressions are introduced, see formulae (\ref{dinclujj}-\ref{dinclugg}).

\item A systematic study is performed, that sheds light on the r\^ole of the 
various model parameters. It is 
emphasized that the forthcoming Tevatron data will allow to constrain their 
values, and answer the 
pending question of the existence of exclusive DPE production. 

\item Precise predictions for discovery physics at 
the LHC will then be possible; we focused on the Higgs boson case.

\item It is  recalled that, assuming favourable exprimental installations, the 
$b\bar{b}$ decay mode of the 
Higgs boson may well be exploited in double diffractive production. 

\item On the theoretical side, the 
H$\rightarrow \tau\tau$ 
channel is most promising, although some experimental difficulties have to be 
overcome.
\end{itemize}

\section*{Acknowledgments}

We wish to thank M. Albrow, A. Bialas, B. Cox, R. Enberg, V. Khoze, R. Orava, A de Roeck, 
M. Ryskin and L. Schoeffel for useful
discussions. One of us (M.B.) is grateful to the CEA/DSM/SPhT for support.

\section*{Appendix}

The cross-sections for the hard processes that are used in our study are 
summarized below, in terms of the usual Mandelstam variables. For dijet
production, we take: 

\begin{eqnarray}
\frac{d\sigma_{gg \rightarrow qq}}{dt} = 
\frac{\pi\alpha_{S}^{2}}{s^{2}}\left(\frac{1}{6}\frac{u^{2}+t^{2}}{ut} 
                              -\frac{3}{8}\frac{u^{2}+t^{2}}{s^{2}}\right) \\
\frac{d\sigma_{gg \rightarrow gg}}{dt} = 
\frac{\pi\alpha_{S}^{2}}{s^{2}}\frac{9}{2}
                               \left(3-\frac{ut}{s^{2}}-\frac{us}{t^{2}}-\frac{st}{u^{2}}\right)
\end{eqnarray}

\noindent For dilepton production, we have:

\begin{eqnarray}
\frac{d\sigma_{qq \rightarrow ll}}{dt} = 
\frac{\pi\alpha^{2}}{s^{2}}\frac{4}{3}\frac{u^{2}+t^{2}}{s^{2}}C_{EW}
\end{eqnarray}

\noindent Here $C_{EW}$ includes electromagnetic and weak factors accounting for
photon and Z exchange between the initial quark pair and the final
leptons. Its expression is, for a quark of flavour $i$:

\begin{eqnarray}
C_{EW} = e_{i}^{2} - e_{i}\frac{s(s-M_{Z}^{2})(L_{e}+R_{e})(L_{q}+R_{q})}                          
{8x_{W}(1-x_{W})[(s-M_{Z}^{2})^{2}+M_{Z}^{2}\Gamma_{Z}^{2}]}
                   + \frac{s^{2}(L_{e}^{2}+R_{e}^{2})(L_{q}^{2}+R_{q}^{2})}
                          {64x_{W}^{2}(1-x_{W})^{2}[(s-M_{Z}^{2})^{2}+M_{Z}^{2}\Gamma_{Z}^{2}]}
\end{eqnarray}

\noindent Above, $x_{W}$ is the weak mixing angle, $M_{Z}$ and $\Gamma_{Z}$ are 
the Z
boson mass and width, and L and R are the left and righthanded chiral couplings.
\noindent Finally, the diphoton production cross-section reads:

\begin{eqnarray}
\frac{d\sigma_{qq \rightarrow \gamma\gamma}}{dt} = 
\frac{\pi\alpha^{2}}{s^{2}}e_{i}^{4}
                                         \frac{1}{3}\left(\frac{u}{t}+\frac{t}{u}\right)\\
\frac{d\sigma_{gg \rightarrow \gamma\gamma}}{dt} = 
\frac{\alpha^{2}\alpha_{S}^{2}}{8\pi s^{2}}
                                         \left(\sum_{i}e_{i}^{2}\right)\frac{1}{8}C_{loop}
\end{eqnarray}

\noindent where $C_{loop}$ is a complicated expression resulting from the
loop integral occurring in the $gg \rightarrow \gamma\gamma$
process. Its detailed expression can be found in \cite{braaten}.

\end{document}